%% file: conference_101719.tex
\documentclass[conference]{IEEEtran}

\IEEEoverridecommandlockouts
\usepackage{cite}
\usepackage{amsmath,amssymb,amsfonts}
\usepackage{algorithmic}
\usepackage{graphicx}
\usepackage{textcomp}
\def\BibTeX{{\rm B\kern-.05em{\sc i\kern-.025em b}\kern-.08em
    T\kern-.1667em\lower.7ex\hbox{E}\kern-.125emX}}
   
\usepackage[table,xcdraw]{xcolor}
\usepackage{multirow}
\usepackage{bigstrut}
\usepackage[ruled,vlined, linesnumbered]{algorithm2e}
\SetKwFor{PostTraverse}{Postorder Traversal}{do}{end}
\SetKw{Continue}{continue}
\SetKw{Break}{break}
\usepackage{blindtext}
\usepackage{mathtools}
\usepackage{amssymb}
\usepackage{tikz}
\usepackage{tikz-cd}
\usepackage{tikzit}
\usepackage[]{amsthm}
\usepackage{float}
\usepackage{listings}
\usepackage{bm}
\usepackage{enumitem}
\usepackage{qcircuit}
\usepackage{physics}
\usepackage{booktabs}
\usepackage{blkarray}
\usepackage[
 	colorlinks=true,
    pdfborder={0 0 0},
    linkcolor=blue
    ]{hyperref}
\input{sample.tikzstyles}

\allowdisplaybreaks

\theoremstyle{definition}

\newtheorem{definition}{Definition}[section]
\newtheorem{example}{Example}[section]
\theoremstyle{plain}
\newtheorem{theorem}{Theorem}[section]

\newtheorem{lemma}[theorem]{Lemma}
\theoremstyle{remark}
\newtheorem{remark}{Remark}[section]
\begin{document}

\title{Efficient CNOT Synthesis for NISQ Devices}

\author{\IEEEauthorblockN{Yao Tang}
\IEEEauthorblockA{\textit{Department of Computer Science} \\
\textit{University of Oxford}\\
Oxford, United Kingdom \\
yao.tang@stx.ox.ac.uk}\\
(Last edited: 12th November 2020)

}

\maketitle

\begin{abstract}
In the era of noisy intermediate-scale quantum (NISQ), executing quantum algorithms on actual quantum devices faces unique challenges. One such challenge is that quantum devices in this era have restricted connectivity: quantum gates are allowed to act only on specific pairs of physical qubits. For this reason, a quantum circuit needs to go through a compiling process called \textit{qubit routing} before it can be executed on a quantum computer. In this study, we propose a CNOT synthesis method called \textit{the token reduction method} to solve this problem. The token reduction method works for all quantum computers whose architecture is represented by connected graphs. A major difference between our method and the existing ones is that our method synthesizes a circuit to an output qubit mapping that might be different from the input qubit mapping. The final mapping for the synthesis is determined dynamically during the synthesis process.  Results showed that our algorithm consistently outperforms the best publicly accessible algorithm for all of the tested quantum architectures.
\end{abstract}


\section{Introduction}
Compositions of CNOT gates on a n-qubit circuit can be represented as $n \times n$ invertible matrices in $\mathbb{F}_2$. Under this representation, a single CNOT gate is simply an elementary matrix, an identity matrix with one off-diagonal entry set to 1. Based on this observation, any parity matrices can be decomposed into a sequence of CNOT gates through Gaussian elimination. Those techniques revealed a new approach for quantum circuit optimization, a task that is critical in the era of NISQ where the feasibility of quantum computing depends heavily on circuit complexity. Additionally, the current physical limits for quantum computers impose a further constraint. Although the theoretical mathematical model for quantum circuits allows interactions between any qubits, the currently available quantum computer architectures only allow specific pairs of qubits to interact. In this study, we present a heuristic method called the token reduction method for synthesis CNOT circuits for any connected architectures. This method not only outperforms the current testable benchmark, but also possess the desired characteristics for routing general quantum circuits.

\subsection{Structure}
In Section \ref{sec: Background}, we introduce the background for the qubit routing problem, and show how CNOT synthesis can be used to solve this. In Section \ref{sec:Preliminaries }, we first show how we can route a circuit to a different output mapping, then we propose a new representation called the \textit{row graph} to work with qubit routing problems. Finally, in Section \ref{sec: Token Reduction Methods} we introduce the token reduction methods.

\section{Background}
\label{sec: Background}

\subsection{The Quantum Circuit Model}

The quantum circuit model is one of the most common ways to represent quantum computation. They are also widely used as models for all quantum processes such as quantum computation, quantum communication, and even quantum noise\cite{10.5555/1972505}. Like classical circuits, quantum circuits consist of wires and logic gates. However, a major departure from
classical logic gates is that the quantum gates describe only unitary linear maps.

Quantum circuits are direct analogs to classical logical circuits, and this means we can interpret a quantum circuit as a “flow” of computation, with the gates acting as operations and the wires indicating the next steps. As a general understanding, a computation starts at the left end of the circuit with some input qubits, and produce the result at the right end of the circuit.

\begin{example}
\label{exp: quantum circuit}
A \textit{Hadamard} gate $H = \input{H.tikz}$\footnote{All graphs in this thesis are generated using TikZiT, available at \href{https://tikzit.github.io/}{https://tikzit.github.io/}} is a linear map:

$$\frac{1}{\sqrt{2}}\begin{pmatrix}
1 & 1\\
1 & -1
\end{pmatrix}$$,

A \textit{CNOT} (Controlled NOT) gate $\input{cnot.tikz}$ is a linear map:

\vspace{12 pt}

$$\begin{pmatrix}
1 & 0 & 0 & 0\\
0 & 1 & 0 & 0\\
0 & 0 & 0 & 1\\
0 & 0 & 1 & 0
\end{pmatrix}$$, 

the black dot is called the \textit{control}, and the white dot with a cross inside is called the \textit{target}. The following circuit computes the unitary linear map:
$$U:=(I \otimes \textbf{H})\circ CNOT \circ(\textbf{H}\otimes I)$$

\begin{figure}[H]
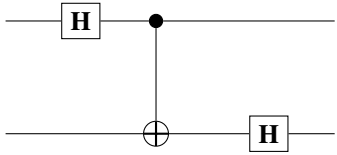

 	\label{fig: quantum_circuit}

	\ctikzfig{quantum_circuit}

	\caption{A simple quantum circuit}
\end{figure}

\end{example}

\subsection{NISQ}
The theories around quantum information and quantum computing are relatively well established, but the quantum computers themselves are still far from ready. In order to run a quantum algorithm such as the Shor's factoring algorithm, it requires a quantum computer that is \textit{fault-tolerant}. Even though quantum technology has advanced rapidly in recent years, we still have a long way to go before achieving fault-tolerant quantum computing. Apart from the qubits needed for computing the algorithms, a fault-tolerant quantum computer also requires millions of qubits as overhead for error correction\cite{O_Gorman_2017}. In comparison, the most recent quantum computer revealed by IBM only has 53 qubits. 

Fault-tolerant quantum computers are still far away, but that is not to say the current or near-future quantum computers are not useful. In fact, some important milestones can potentially be reached by these devices such as the first demonstration of quantum computers solving a problem that is infeasible for classical computers \footnote{A recent work claimed that this milestone had been achieved; however, whether the claim is true is still in debate\cite{supremacy}}. Preskill\cite{Preskill_2018} called these quantum computers, the \textit{noisy intermediate-scale quantum (NISQ) devices}. 

Other than the most noticeable issue of not having quantum error correction (i.e., not fault-tolerant), these NISQ devices also have other limitations. Firstly, the error rate for quantum gates are significantly high. The overall fidelity of computations on NISQ devices is directly affected by the depth of the circuits; the computations get noisier as the circuits get deeper \cite{Preskill_2018}. 

Secondly, these devices have \textit{restricted connectivity} such that any two-qubit quantum gate must satisfy certain topological restrictions; this usually means that a two-qubit gate can only be applied to a pair of adjacent qubits and with a certain direction. 

\begin{figure*}[!t]
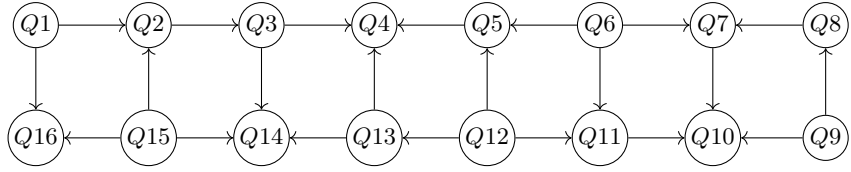

 	\label{fig: ibm_qx5}
	\ctikzfig{ibm_qx5}
	\caption{IBM QX5 Architecture}
\end{figure*}

Figure \ref{fig: ibm_qx5} shows the IBM QX5 architecture, which uses CNOT gates as the only allowed two-qubit gates; the edges indicate the connectivity between qubits, and the arrows indicate the allowed directions for CNOT gates. For example, the pair $(Q1, Q2)$ only allows CNOT gates whose control acts on $Q1$ and whose target acts on $Q2$.

\subsection{Qubit Routing}
\label{sec: Qubit Routing Problem}
We mentioned that the NISQ devices have restricted connectivity:
\begin{enumerate}
	\item A two-qubit gate can only be applied to adjacent qubits.
	\item A two-qubit gate can only be applied to adjacent qubits with certain directions.
\end{enumerate}

In this paper, however, we will only consider the first constraint. The second constraint is specific to CNOT gates, and it can usually be solved by the Equation (\ref{eq: cnot direction}) after the first constraint is met through compilation. Nevertheless, the gate direction constraint can still be an important factor to consider when we design methods to solve the first constraint; we will discuss this in Section \ref{sec: Future Work}. As a result, we will define quantum architectures as \textit{undirected} graphs in Definition \ref{def: architecture graph}, instead of the one shown in Figure \ref{fig: ibm_qx5}.

\begin{equation}
\label{eq: cnot direction}
\input{cnot.tikz} = \input{cnot_reverse.tikz} 
\end{equation}\\

Before we demonstrate why the restricted connectivity can be an issue, let's explain how a quantum circuit can be applied to a quantum computer. Firstly, let's define a quantum computer architecture in terms of graphs. 

\begin{definition}[Architecture Graph]
\label{def: architecture graph}
An architecture graph is an \textit{undirected} connected graph $G\langle V, E \rangle$, such that each node represents a qubit, and each edge represents a connection. 
\end{definition}

\begin{remark}
The usage of the term qubit is sometimes confusing. In this thesis, we use the term \textit{qubit} or \textit{node} for the physical nodes on quantum computers. We use \textit{logical qubit} or \textit{wire} for the wires in quantum circuits.
\end{remark}

A quantum circuit must be mapped to a an architecture graph before it can be executed. Therefore, we define the following:

\begin{definition}[Mapping]
Given a quantum circuit with $n$ wires $W=\{w_1, w_2 \cdots w_n\}$, and an architecture graph $G\langle V, E \rangle$ with $|V|=n$,  a \textit{mapping} is a bijective function: 
$$f: W \to V$$
\end{definition}

In the following example, we illustrate how a quantum computer computes a circuit.
\begin{example}
\label{exp: run}
Consider the following architecture graph $G$, the mapping $M$ and circuit $c$. 
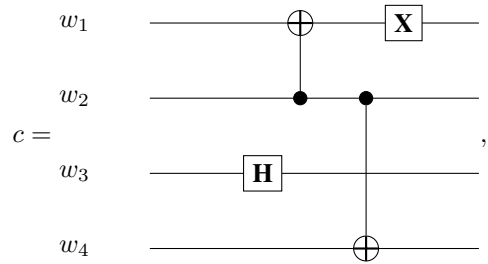
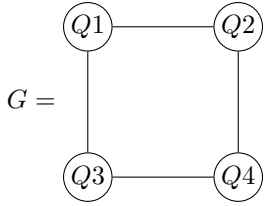
$$c = \input{ex_routing_1_c.tikz}, $$
$$G = \input{ex_routing_1_g.tikz} $$

$$M=\begin{cases}
w_1 \mapsto Q1 \\
w_2 \mapsto Q2 \\
w_3 \mapsto Q3 \\
w_4 \mapsto Q4 \\
\end{cases}$$

When given a set of inputs, the quantum device whose architecture is represented by $G$ computes $c$ with mapping $M$ as follows:

\begin{figure}[H]
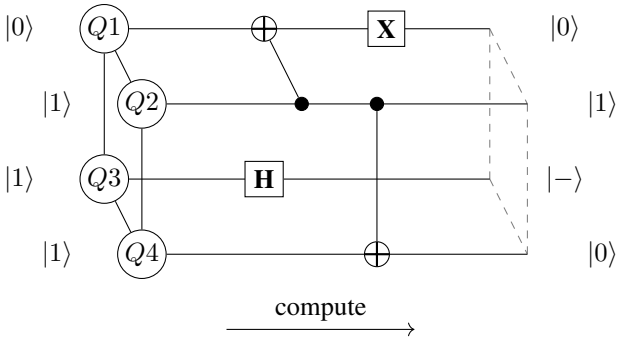

 	\label{fig: ex_routing_1_1_two}
	\ctikzfig{ex_routing_1_1_two}
	\caption{Mapping without routing}
\end{figure}

$\input{X.tikz}$ is the Pauli X gate, whose matrix is $\begin{pmatrix}
0 & 1\\
1 & 0\\
\end{pmatrix}$
\end{example}

\begin{figure}[H]
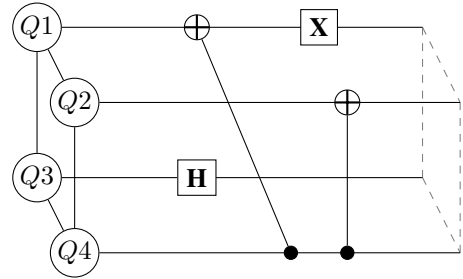

 	\label{fig: ex_routing_1_3}
	\ctikzfig{ex_routing_1_3}
	\caption{Mapping requires routing}
\end{figure}

In Example \ref{exp: run}, the architecture $G$ computes the circuit $c$ with mapping $M$ without any issue. However, $G$ cannot compute with the following mapping $M'$ as shown in Figure \ref{fig: ex_routing_1_3}. The first CNOT gate acts on the qubits $Q1$, $Q4$, which are not adjacent.

$$M'=\begin{cases}
w_1 \mapsto Q1 \\
w_2 \mapsto Q4 \\
w_3 \mapsto Q3 \\
w_4 \mapsto Q2 \\
\end{cases}$$

However, this circuit can still compile with mapping $M'$ by inserting SWAP gates. A SWAP gate is a linear map that swaps qubits between two wires. 

\begin{equation}
\label{eq:SWAP EQ}
SWAP=\input{SWAP.tikz} \cong  \begin{pmatrix}
1 & 0 & 0 & 0\\
0 & 0 & 1 & 0\\
0 & 1 & 0 & 0\\
0 & 0 & 0 & 1
\end{pmatrix}
\end{equation}

\begin{figure}[H]
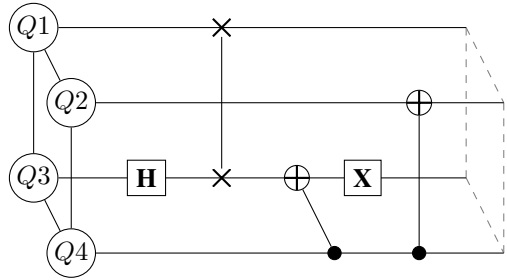

 	\label{fig: ex_routing_1_routed}
	\ctikzfig{ex_routing_1_routed}
	\caption{Routed circuit $c'$}
\end{figure}

\begin{remark}
The circuit $c$ shown in Example \ref{exp: run} can always be compiled with architecture $G$ with an appropriate initial mapping without using any routing. However, this is not always possible as the circuit can get very large. Suppose another CNOT gate acting on $(w_1, w_4)$ is added to $c$, then one can verify that no initial mapping is sufficient for the circuit to compile with $G$ if no extra routing is performed. Nevertheless, selecting a optimal initial mapping that minimizes the number of SWAP gates is an NP-hard problem \cite{inproceedings}, and it is shown that the gate count in a routed circuit is very sensitive to the initial mapping. Hence optimizing initial mapping is often a very important task for any routing algorithms. 
\end{remark}

An important observation is that the circuit $c'$ and our original circuit $c$ do not compute the exact same function. In fact, the SWAP gate only changed the mapping of the output, and the output mapping is:

$$M'_t=\begin{cases}
w_3 \mapsto Q1 \\
w_2 \mapsto Q4 \\
w_1 \mapsto Q3 \\
w_4 \mapsto Q2 \\
\end{cases}$$

\vspace{12 pt}

This change of mapping can be understood as the computation result being output with a permutation that differs from the input's permutation. This is a totally acceptable behavior  because quantum gates are expensive, and we are better off re-ordering the results in a classical way after the quantum computation. 

Since a routed circuit is still equivalent to the original circuit if only the output mappings are different, we define quantum circuit routing procedures as the following.

\begin{definition}
A circuit routing procedure is defined as :\\
Given a $n$-wire quantum circuit $c$, and a $n$-qubit architecture graph $G$. Outputs a $n$-qubit circuit $c'$, an initial mapping $M_0$ and a output mapping $M_t$ satisfying:
\begin{enumerate}
	\item $c'$ complies with $G$ with mapping $M_0$.
	\item $c'$ is equivalent to $c$ with the results mapped by $M_t$.
\end{enumerate}
\end{definition}

Since the fidelity of computation is critically affected by the depth of the circuits in the NISQ era, the most crucial goal for a routing procedure is to minimize the number of gates. 

\subsection{CNOT Synthesis}

A universal gate set for quantum computing often contains one choice of two-qubit gates and some single-qubit gates. Usually, CNOT gates are used as the choice for primitive two-qubit gates. For example, in IBM Q devices, CNOT gates are the only allowed two-qubit gates; all other two-qubit gates need to be converted to CNOTs before applying. In particular, SWAP gates and CZ gates can be converted to CNOT gates by the following equations:

\begin{equation}
CZ = \input{cz.tikz} = \input{CZ_3.tikz}
\end{equation} 

\begin{align}
\label{eq: swap to cnot}
SWAP = \input{SWAP.tikz} = \input{SWAP_CNOT.tikz} \nonumber \\
=\input{SWAP_CNOT2.tikz}
\end{align} 

Treating CNOT as the only two-qubit gates not only simplifies the qubit routing problem as we can partition the two-qubit gates into blocks of CNOT sub-circuits,  but also opened a entirely new perspective. CNOT gates alone form a very important subclass of quantum circuits called the linear reversible classical circuits\cite{Patel}. This type of circuits implements all linear reversible functions $f: \mathbb{F}_2^n \to \mathbb{F}_2^n$\footnote{Here $\mathbb{F}_2$ is the Galois field of two elements \{0,1\}}, where $n$ is the number of wires. In other words, all CNOT circuits with $n$ wires can be represented as an invertible $n \times n$ matrix in $\mathbb{F}_2$. This representation allows us to generate a CNOT circuit from a matrix using matrix decomposition methods such as Gaussian eliminations. We call this gate generation procedure \textit{CNOT synthesis}.

A CNOT gate in classical circuits has the following truth table:
\begin{figure}[H]
 	\label{fig: CNOT truth table}
	\begin{displaymath}
	\begin{array}{|c c|c c|}
	% |c c|c| means that there are three columns in the table and
	% a vertical bar ’|’ will be printed on the left and right borders,
	% and between the second and the third columns.
	% The letter ’c’ means the value will be centered within the column,
	% letter ’l’, left-aligned, and ’r’, right-aligned.
	\multicolumn{2}{c}{Input} & \multicolumn{2}{c}{Output}\\
	\hline 
	control & target & control & target\\ % Use & to separate the columns
	\hline % Put a horizontal line between the table header and the rest.
	0 & 0 & 0 & 0\\
	0 & 1 & 0 & 1\\
	1 & 0 & 1 & 1\\
	1 & 1 & 1 & 0\\
	\end{array}
	\end{displaymath} 
	\caption{CNOT Truth Table}
\end{figure}

Considering CNOT gates as classical linear reversible operators gives us a different matrix representation than the one shown in Example \ref{exp: quantum circuit}. In this new representation, CNOT gates are surprisingly just \textit{elementary matrices}. That is, an $n \times n$ identity matrix with an additional off-diagonal entry set to 1. For example:

$$\begin{pmatrix}
1 & 0 & 0\\
0 & 1 & 0\\
1 & 0 & 1\\
\end{pmatrix} \sim  \input{CNOT_2.tikz}$$

\vspace{12 pt}

\begin{theorem}
Given an $n$-wire CNOT circuit with wires ordered by $[0,1,\cdots, n-1]$, a CNOT gate whose control placed on the $c_{th}$ wire and target placed on the $t_{th}$ wire is represented by a $n \times n$ identity matrix with an additional off-diagonal 1 in the $(t,c)$ entry, that is, the $c_{th}$ column and $t_{th}$ row.  
\end{theorem}

Therefore, we can treat each CNOT gate as an elementary row operation.

\begin{theorem}
\label{thm: row operation}
The action of a CNOT gate, whose control is on the $c_{th}$ wire and target is in the $t_{th}$ wire, is an \textit{elementary row operation} on a $n \times n$ matrix on $\mathbb{F}_2$, which adds the $c_{th}$ row to the $t_{th}$ row.

\end{theorem}

The following two theorems serve as the foundation for the concept of CNOT synthesis. 
\begin{theorem}
\label{thm: cnot synthesis}
Any classical linear reversible circuit can be synthesized by a sequence of CNOT gates.
	\begin{proof}
		For a classical linear reversible circuit with $n$ wires, suppose it is represented by a $n\times n$ matrix $A$ in $\mathbb{F}_2$. Then by the fact that any invertible matrix can be reduced to an identity matrix by using row operations, $E_{k}E_{k-1}\cdots E_{1}A=I  $ where $E_{j}$ is an elementary matrix. Hence $A=E_{1}^{-1}\cdots E_{k-1}^{-1}E_{k}^{-1}$. But since these elementary matrices are self-inverse in $\mathbb{F}_2$ (i.e.$E^{-1} = E$ ), we obtain $A = E_{1}\cdots E_{k-1}E_{k}$. 
		
		That is, we construct a CNOT circuit from $A$ by adding the CNOT gates corresponding to $E_{1}\cdots E_{k-1}E_{k}$ in reverse order.
	\end{proof}
\end{theorem}

\begin{remark}
The fact that elementary matrices in $\mathbb{F}_2$ are self-inverse is also an important aspect for quantum circuit optimization as gate cancellation is often the easiest way to reduce gate count. The reduction is achieved by the identity shown in Equation (\ref{eq: cnot identity}).

\begin{equation}
\label{eq: cnot identity}
\input{double_cnot.tikz}= \input{Identity.tikz}\\[12 pt]
\end{equation}  
\end{remark}

\begin{theorem}
\label{thm: upper bound}
The upper bound for the number of CNOTs needed to synthesis any $n$-wire classical linear reversible circuit is $\Omega(n^2/\log(n))$ \cite{Patel}.
\begin{proof}
	For a given linear reversible circuit with $n$ wires, let $d$ be the number of CNOT gates needed for synthesis the circuit. We know that there are $n(n-1)$ different placements for each CNOT, hence there are in total $(n^2-n)^d$ different CNOT circuits. Because there are $O(2^{n^2})$ different linear maps for the original circuit (i.e. $2^{n^2}$ different $n\times n $ matrices), we need $(n^2-n)^d \geq 2^{n^2}$. Take $\log$ for both sides we have $d \log(n^2-n) \geq n^2$ hence $d \geq n^2/\log(n^2-n) =  \Omega(n^2/\log(n))$.
\end{proof}
\end{theorem}

Theorem \ref{thm: cnot synthesis} demonstrates how we can synthesis a CNOT circuit from any classical linear reversible circuit. At this point, these theorems at least give us an idea of how a given CNOT circuit can be potentially optimized through synthesis. If a given CNOT circuit has a very large number of gates, there is a good chance that there is an equivalent but smaller circuit. 

The methods for CNOT synthesis without connectivity constraints are well researched. Patel et al. \cite{Patel} show that the upper bound in Theorem \ref{thm: upper bound} can be achieved through an optimized version of LU decomposition. A standard framework for how synthesis work is illustrated in Algorithm \ref{algo: standard synthesis}. 

\begin{algorithm}
\label{algo: standard synthesis}
	\SetAlgoLined
	\KwIn{Classical reversible circuit $c$ }
	Construct an empty circuit $c'$\;
	Construct a matrix $P$ from $c$ by composing the matrices representing the gates.\;
	Perform matrix decomposition on $P$ using only row operations until an identity matrix is reached. Add the CNOT gates corresponding to each row operation to $c'$ in reverse order (i.e. from right to left)\;
	\Return{$c'$}
	 \caption{\textsc{Template for CNOT Synthesis}}
\end{algorithm}

Theorem \ref{thm: row operation} states that composing a CNOT gate to a classical linear reversible circuit is equivalent to performing an elementary row operation to the matrix representation of the circuit. When connectivity is an issue, however, it poses a constraint on which row operations are allowed. We need to decompose a matrix to identity using only allowed row operations implied by an architecture graph (Definition \ref{def: architecture graph}); otherwise, the resulting CNOT circuit may not be compatible with the architecture due to invalid connectivities.

\begin{definition}[Constrained CNOT synthesis]
A \textit{constraint CNOT synthesis} procedure is defined as :\\
Given a $n$-wire CNOT circuit $c$, and a $n$-qubit architecture graph $G$. Outputs a $n$-qubit circuit $c'$, an initial mapping $M_0$ and a output mapping $M_t$ satisfying:
\begin{enumerate}
	\item $c'$ complies with $G$ with mapping $M_0$.
	\item $c'$ is equivalent to $c$ with the results mapped by $M_t$.
\end{enumerate}
\end{definition}

Two recent independent works \cite{kissinger2019cnot}\cite{nash2019quantum} efficiently solve the constrained CNOT synthesis problem using a method called Steiner Gauss. Roughly speaking, they innovate to use Steiner trees (Definition \ref{def: steiner tree}) to perform LU decomposition with only the row operations specified by a target architecture. Steiner Gauss achieves a $\Omega(n^2)$ upper bound for the number of CNOT gates generated and runs efficiently fast (i.e., $O(n^3)$). Noticeably, the work done by Kissinger and Van de Griend shows a significant improvement in gate count comparing to two SWAP insertion based techniques \cite{smith2016practical} \cite{alex2019qubit}, in some of the test circuits, the Steiner Gauss method results in more than 80\% savings against the other two techniques. This discovery opens a new door for using CNOT synthesis as a way to solve the qubit routing problem. 

\begin{example}
The following circuit:

\begin{figure}[H]
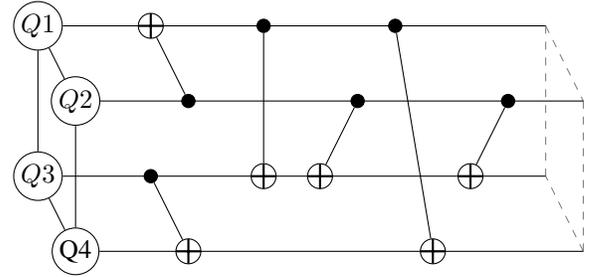

	\ctikzfig{case_reduced}
	\caption{Original circuit} \label{fig: case_reduced}
\end{figure}

has a synthesised circuit

\begin{figure}[H]
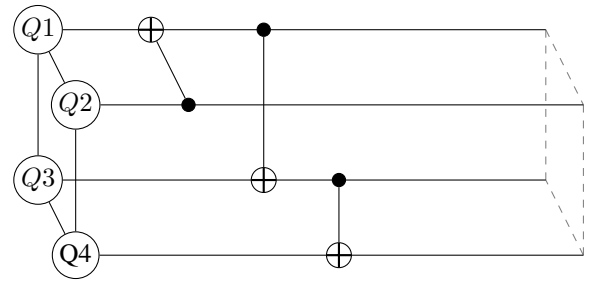

	\ctikzfig{case_reduced_r}
	\caption{Synthesized circuit} \label{fig: case_reduced_r}
\end{figure}
\end{example}

\begin{definition}[Unweighted Rooted Steiner Tree]
\label{def: steiner tree}
Let $G = \langle V, E \rangle$ be an unweighted, undirected graph, and let $S \subseteq V$ be a subset of vertices containing a root node $u$, called the \textit{terminals}. A Steiner tree is a tree $\langle V_T, E_E \rangle $ in $G$ that spans $S$ with root $u$. We also call the set of vertices $V_T \setminus V$ the \textit{Steiner points}.
\end{definition}

\begin{remark}
Finding the optimal Steiner Tree is an NP-complete problem, we give an simple heuristic algorithm that uses the Floyd-Warshall algorithm in Appendix \ref{app:steiner}.
\end{remark}

\section{Preliminaries}
\label{sec:Preliminaries }
The Steiner Gauss method works great for CNOT only quantum circuit; however, it does not work very well if applied naively to a general circuit. 

Let's consider a circuit with only a single CNOT $g$. Inserting SWAP gates is essentially trying to compute $g$ to a different output mapping. However in many cases, a circuit synthesized using the Steiner Gauss method for $g$ resembles a circuit that firstly computes $g$ to a different output mapping and then reverts the output to the original mapping (see Example \ref{exp: intuition}). Therefore, if CNOT gates are sparse, which is common for quantum circuits, any synthesis methods whose result has the same input and output mapping could perform too many unnecessary mapping reversions hence increasing gate count. 

CNOT synthesis has the potential benefit of computing an arbitrary function $f$ with a small set of CNOTs; this is why when a CNOT block is large, the Steiner Gauss method performs better even though it still has unnecessary mapping reversion. However, when the block is small, synthesis without change mappings is unlikely to generate a smaller set of CNOTs.

In this study, we introduce an efficient CNOT synthesis method that produces a routed circuit according to an architecture that is equivalent to the original circuit up to an output mapping.
\begin{example}
\label{exp: intuition}
Consider the following mapped circuit.
 	\begin{figure}[H]
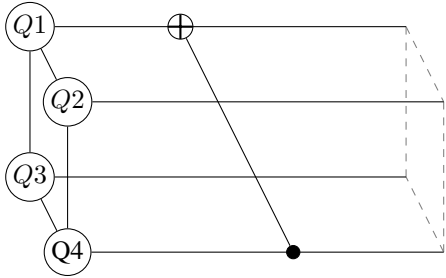

		\ctikzfig{synthesis}
		\caption{Original circuit} \label{fig: synthesis}
	\end{figure}
Routing the circuit through SWAP insertion can give us the following (recall from (\ref{eq: swap to cnot}), a SWAP gate can be converted to three consecutive CNOT gates.):
 	\begin{figure}[H]
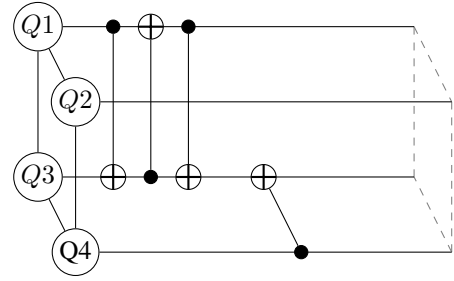

		\ctikzfig{synthesis_swap}
		\caption{Routing with SWAP} \label{fig: synthesis_swap}
	\end{figure}
The CNOT synthesis method can give us the following:
 	\begin{figure}[H]
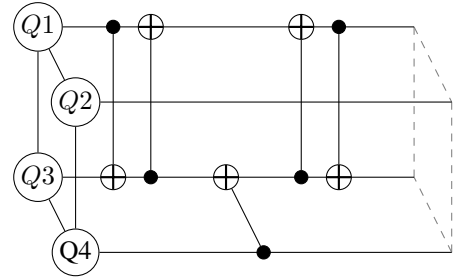

		\ctikzfig{synthesis_cnot}
		\caption{Routing with CNOT synthesis} \label{fig: synthesis_cnot}
	\end{figure}
but notice, this circuit is equivalent to the following based on the identity (\ref{eq: cnot identity}):
 	\begin{figure}[H]
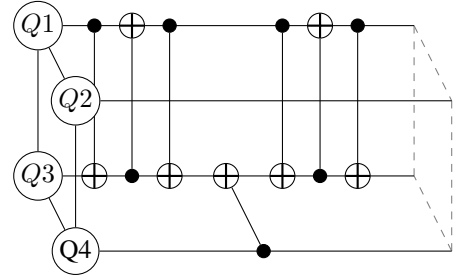

		\ctikzfig{synthesis_eqiuvalency}
		\caption{Routing with CNOT synthesis (2)} \label{fig: synthesis_eqiuvalency}
	\end{figure}
\end{example}

\subsection{Matrix Decomposition to a Permutation Matrix}

\begin{theorem}
\label{thm: compose with permutation matrix}
A CNOT circuit can be synthesized to a different mapping by left-multiply its associated matrix with the transpose of a permutation matrix and then perform the synthesis.
\end{theorem}

\begin{definition}[Permutation Matrix]
A permutation matrix is a square matrix obtained by permuting the \textit{columns} of an $n \times n$ identity matrix according to some permutation $\pi$. 
\end{definition}

\begin{example}
Suppose there is a $n\times n$ matrix $P = \begin{pmatrix}
	  p_{1,1} & p_{1,2} & \cdots & p_{1,n} \\
	  p_{2,1} & p_{2,2} & \cdots & p_{2,n}  \\
	  \vdots & \vdots & \cdots & \vdots &  \\
	  p_{n,1} & p_{n,2} & \cdots & p_{n,n}  \\
	  \end{pmatrix}$ . We want to transform $P$ to $P'$ such that:
	  
	$$P' \cdot \begin{pmatrix}
	  x_0 \\
	  x_1  \\
	  \vdots \\
	  x_{n-1}
	\end{pmatrix} = \begin{pmatrix}
	  p_{2,0}x_0+p_{2,0}x_1+\cdots+p_{2,0}x_n \\
	  p_{n,1}x_1+p_{n,2}x_2+\cdots+p_{n,n}x_n \\
	  \vdots \\
	  p_{1,1}x_1+p_{1,2}x_2+\cdots+p_{1,n}x_n \\
	\end{pmatrix}$$
	
	\vspace{12 pt}
	
	That is, we want to change the output permutation from [1 2 3 ... n] to [2 n 3 ... 1]. This change of permutation is represented uniquely by the following permutation matrix:
	
	$$\textbf{m} =\begin{pmatrix}
	  0 & 0 & \cdots & 1 \\
	  1 & 0 & \cdots  & 0  \\
	  \vdots & \vdots & \cdots  & \vdots  \\
	  0 & 1 & \cdots & 0 
	\end{pmatrix}$$
	
	Notice that $ P'=\textbf{m}^{T} \cdot P$
\end{example}

This utilization of permutation matrices leads to a more general routing procedure shown as Algorithm \ref{alg:General Qubit Routing}.

	\begin{algorithm}
	\label{alg:General Qubit Routing}
	\SetAlgoLined
	\KwIn{A quantum circuit $c$ }
	\KwOut{A quantum circuit $c'$, an initial mapping $M_0$, and output mapping $M_t$}
	Optimize an initial mapping $M_0$\;
	Initialize $c'$ to an empty circuit\;
	Partition $c'$ into blocks of gates such that each block contains only CNOT gates or one-qubit gates\;
	$M_t \gets M_0$\;
	\While{next block of gates $\mathcal{B}$ in $c'$}{
		\eIf{$\mathcal{B}$ fits $M_t$ }{
			add $\mathcal{B}$ to $c'$\;
			\Continue		
		}{
			Find an optimal mapping $M'$\;
			Synthesize $\mathcal{B}$ to $\mathcal{B}'$ that has a output mapping $M'$\;
			$M_t \gets M'$
			Add $\mathcal{B}'$ to $c'$\;
		}
	}
	\Return{$c', M_0, M_t$}
	 \caption{\textsc{General qubit routing with CNOT synthesis}}
	\end{algorithm}

\vspace{12 pt}

However, this procedure still has one problem: finding an optimal output mapping is difficult. The output mapping needs to be optimized such that:
\begin{enumerate}
\item The matrix is easy to synthesize.
\item The output mapping should also benefit the rest of the circuit. (The output mapping $M_t$ for the current CNOT block is optimal in a sense that the rest of the circuit $c'$ requires the minimal number of gates to be routed given $M_t$ as the initial mapping for $c'$.)
\end{enumerate}

Finding such mapping has a $n!$ search space. We want a synthesis procedure to decide an output permutation \textit{on the fly}. The following theorem shows that it is possible.

\begin{theorem}[Decomposition to a permutation matrix]
\label{thm: decompose to permutation}
Suppose $P$ is the matrix representing a CNOT circuit with respect to the original permutation. Let $\textbf{m}$ be a permutation matrix associated with some permutation $\pi$. We can synthesize $P$ to the permutation $\pi$ by decomposing $P^{T}$ to $\textbf{m}^{T}$.

\begin{proof}
If we synthesize $P^{T}$ to $\textbf{m}^{T}$ then,
\begin{align*}
P^{T}&=E_1\cdot E_2\cdot \cdots \cdot E_n \cdot \textbf{m}^{T}\\
&\implies P^{T}\cdot \textbf{m} = E_1\cdot E_2\cdot \cdots \cdot E_n && (\textbf{m} \text{ is orthogonal})\\
&\implies \textbf{m}^{T} \cdot P = E_n^{T}\cdot E_{n-1}^{T}\cdot \cdots \cdot E_1^{T}
\end{align*}

\vspace{12 pt}

That is, we effectively obtain a decomposition of $\textbf{m}^{T} \cdot P$ (see Theorem \ref{thm: compose with permutation matrix}) by decomposing $P^{T}$ to $\textbf{m}^{T}$. Then, we can add the CNOT gates associated with $E_1^{T}, E_2^{T},....,E_n^{T}$ one by one to the circuit. Notice that the transpose of of a CNOT matrix represents a CNOT gate with the control and target switched.
\end{proof}
\end{theorem}

Next, we give an example of how above theorem works.

\begin{example}
\label{exp: permuted synthesis}
Consider the following circuit $c$ with a mapping $M$ to a architecture $G$
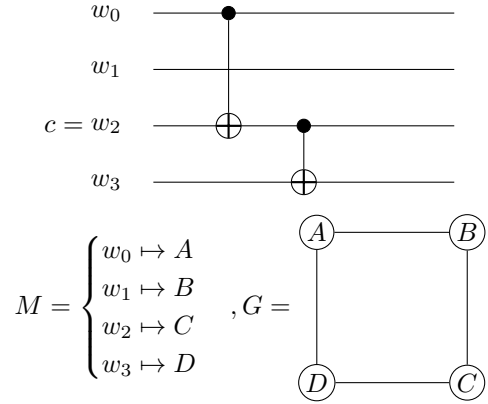
$$c=\input{synthesis_permutation_c.tikz}$$
$$M=\begin{cases}
	w_0 \mapsto A\\
	w_1 \mapsto B\\
	w_2 \mapsto C\\
	w_3 \mapsto D
\end{cases}, G = \input{4_grid.tikz}$$
Hence the circuit $c$ corresponds to the following permutated matrix $P$

$$P=\begin{blockarray}{cccc}
	\begin{block}{(cccc)}
	  1 & 0 & 0 & 0 \\
	  0 & 1 & 0 & 0  \\
	  0 & 0 & 1 & 0  \\
	  0 & 0 & 1 & 1 \\
	\end{block}\end{blockarray} \cdot 
\begin{blockarray}{cccc}
	\begin{block}{(cccc)}
	  1 & 0 & 0 & 0 \\
	  0 & 1 & 0 & 0 \\
	  1 & 0 & 1 & 0\\
	  0 & 0 & 0 & 1\\
	\end{block}\end{blockarray} 
=\begin{blockarray}{ccccc}
	\begin{block}{(cccc)c}
	  1 & 0 & 0 & 0 & A\\
	  0 & 1 & 0 & 0 & B \\
	  1 & 0 & 1 & 0 & C \\
	  1 & 0 & 1 & 1 & D \\
	\end{block}
\end{blockarray}  $$

%$$P=\begin{pmatrix}
%	  1 & 0 & 0 & 0 \\
%	  0 & 1 & 0 & 0  \\
%	  0 & 0 & 1 & 0  \\
%	  0 & 0 & 1 & 1 \\
%\end{pmatrix} \cdot \begin{pmatrix}
%	  1 & 0 & 0 & 0 \\
%	  0 & 1 & 0 & 0 \\
%	  1 & 0 & 1 & 0\\
%	  0 & 0 & 0 & 1\\
%\end{pmatrix} =\begin{blockarray}{ccccc}
%	\begin{block}{(cccc)c}
%	  1 & 0 & 0 & 0 & A\\
%	  0 & 1 & 0 & 0 & B \\
%	  1 & 0 & 1 & 0 & C \\
%	  1 & 0 & 1 & 1 & D \\
%	\end{block}
%\end{blockarray}  $$.
And 
$$P^{T} = \begin{blockarray}{ccccc}
	\begin{block}{(cccc)c}
	  1 & 0 & 1 & 1 & A\\
	  0 & 1 & 0 & 0 & B \\
	  0 & 0 & 1 & 1 & C \\
	  0 & 0 & 0 & 1 & D \\
	\end{block}
\end{blockarray}$$
The objective is to decompose $P^T$ to some transposed permutation matrix $\textbf{m}^T$ with only row operations allowed by the connectivity in $G$. In order to work on this task, it is visually easier if we represent $P^T$ and $G$ together such that each row of $P^T$ is placed on a node in $G$. 

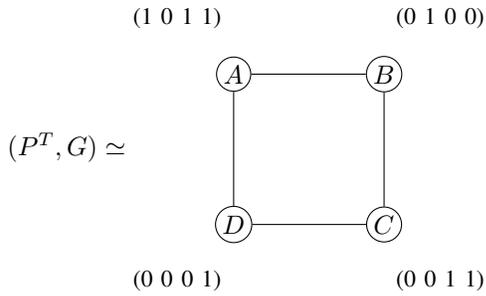
\begin{equation}
\label{eq: row graph}
(P^{T}, G) \simeq \input{matrix_graph.tikz} 
\end{equation}

First, notice that we can swap the row vectors on A and B such that $\begin{pmatrix}1 & 0 & 1 & 1\end{pmatrix}$ can be transformed to $\begin{pmatrix}1 & 0 & 0 & 0\end{pmatrix}$ by the row operation $Row(B) = Row(B)+Row(C)$. Lastly, the vector on $C$ can be transformed to $\begin{pmatrix}1 & 0 & 1 & 0\end{pmatrix}$ by $Row(C) = Row(C)+Row(D)$

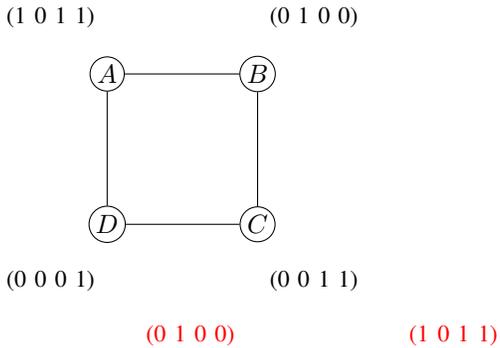
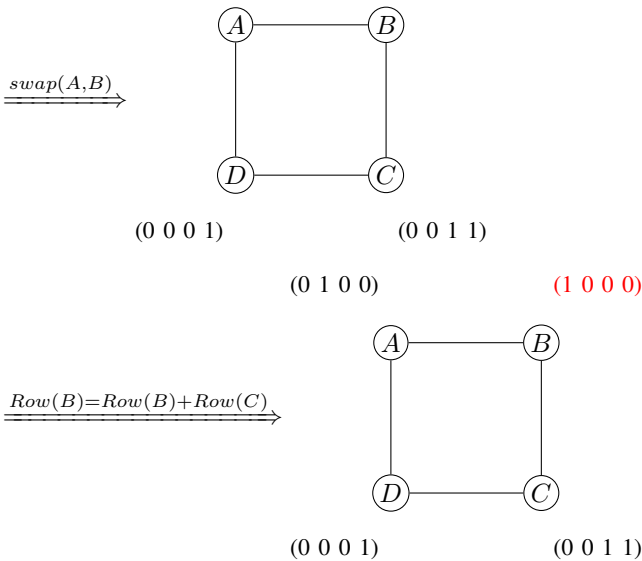
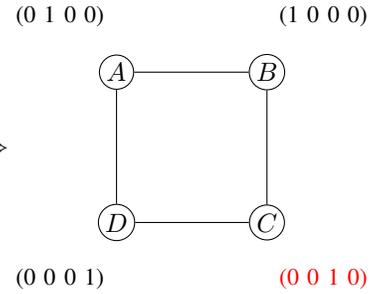
\begin{align*}
&\input{matrix_graph.tikz} \\[10 pt]
&\xRightarrow[]{swap(A,B)} \input{matrix_graph_swap.tikz}\\[10 pt]
&\xRightarrow[]{Row(B) = Row(B)+Row(C)} \input{matrix_graph_B_C.tikz}\\[10 pt]
&\xRightarrow[]{Row(C) = Row(C)+Row(D)} \input{matrix_graph_C_D.tikz}
\end{align*}

Recall a swap is simply three consecutive row operations in $\mathbb{F}_2$ (i.e. Equation (\ref{eq: swap to cnot})). Hence $P^{T}$ has the following decomposition:

\begin{align*}
P^{T} &=\begin{pmatrix}
	  1 & 1 & 0 & 0 \\
	  0 & 1 & 0 & 0  \\
	  0 & 0 & 1 & 0  \\
	  0 & 0 & 0 & 1 \\
\end{pmatrix} \cdot \begin{pmatrix}
	  1 & 0 & 0 & 0 \\
	  1 & 1 & 0 & 0  \\
	  0 & 0 & 1 & 0  \\
	  0 & 0 & 0 & 1 \\
\end{pmatrix} \cdot \begin{pmatrix}
	  1 & 1 & 0 & 0 \\
	  0 & 1 & 0 & 0  \\
	  0 & 0 & 1 & 0  \\
	  0 & 0 & 0 & 1 \\
\end{pmatrix}\\[12 pt]
&\cdot \begin{pmatrix}
	  1 & 0 & 0 & 0 \\
	  0 & 1 & 1 & 0  \\
	  0 & 0 & 1 & 0  \\
	  0 & 0 & 0 & 1 \\
\end{pmatrix}\cdot \begin{pmatrix}
	  1 & 0 & 0 & 0 \\
	  0 & 1 & 0 & 0  \\
	  0 & 0 & 1 & 1  \\
	  0 & 0 & 0 & 1 \\
\end{pmatrix}\cdot \begin{pmatrix}
	  0 & 1 & 0 & 0 \\
	  1 & 0 & 0 & 0  \\
	  0 & 0 & 1 & 0  \\
	  0 & 0 & 0 & 1 \\
\end{pmatrix}
\end{align*}

and $\textbf{m}^T = \begin{pmatrix}
	  0 & 1 & 0 & 0 \\
	  1 & 0 & 0 & 0  \\
	  0 & 0 & 1 & 0  \\
	  0 & 0 & 0 & 1 \\
\end{pmatrix}$.

Finally, we construct a new circuit $c'$ following the procedure in Theorem \ref{thm: decompose to permutation}.
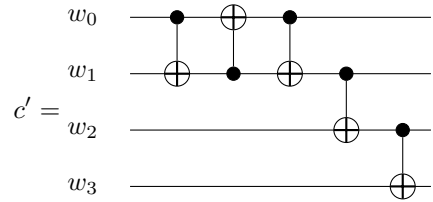
$$c'=\input{synthesis_permutation_c_p.tikz}$$
Observe the the permutation of the output mapping is indicated by the permutation matrix $\textbf{m}$.
\end{example}

In the above example we assign each row vector in a matrix to its corresponding node to have a more visually intuitive model to work with. In the next section, we formally define this representation and replace row vectors by the sums of the standard basis vectors.

\subsection{Row Graph}

\begin{definition}[Row Graph]
\label{def: Row Graph}
A row graph is a triple $(G,P,f)$, where
\begin{enumerate}
\item $P$ is a $n \times n$ matrix
\item $G \langle V,E \rangle$ is a connected undirected graph such that $|V| = n$.
\item $f: V \to P_{rows}$ is an assignment function, where $P_{rows}$ is the set of row vectors in matrix $P$.
\end{enumerate}
Furthermore, each row vector is represented as a sum of the standard basis vectors. For example, $\begin{pmatrix}1 & 0 & 1 & 0\end{pmatrix}$ is represented as $e_0 + e_2$, where $e_i$ is the standard basis vector in $\mathbb{F}_2^n$ with $i_{th}$ entry set to 1 and 0 elsewhere. We also denote $size(f(u))$ to be the number of ones in a row vector $f(u)$.
\end{definition}

\begin{example}
The following is a row graph where $V=\{A,B,C,D\}$, $P=\begin{pmatrix}
	  1 & 1 & 0 & 0 \\
	  1 & 0 & 0 & 0  \\
	  1 & 0 & 1 & 1  \\
	  0 & 1 & 0 & 0 \\
\end{pmatrix}$.
	\begin{figure}[H]
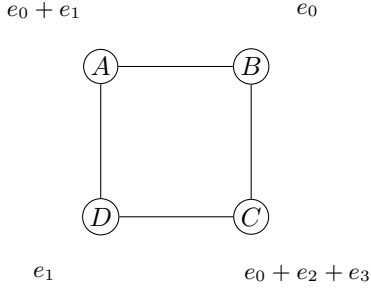

		\ctikzfig{row_graph}
		\caption{Row graph example} \label{fig: row_graph}
	\end{figure}
\end{example}

Moreover, we rename the elementary row addition operation as the node addition operation to suit our new representation.

\begin{definition}[Node addition operation]
Given two nodes $u,v$ on a node graph and an assignment function $f$, we define the node addition operation as $f(u) \gets f(u) + f(v) $. Recall this is the addition in $\mathbb{F}_2$.
\end{definition}

\begin{remark}[Swap on row graphs]
A swap operation is also an very important operation on row graphs. For any two nodes $u,v$, $Swap(u,a)$ can be performed by $f(u) \gets f(u) + f(v)$,  $f(v) \gets f(u) + f(v)$ and $f(u) \gets f(u) + f(v)$.  $swap(u,v)$ can be also performed by $f(v) \gets f(u) + f(v)$,  $f(u) \gets f(u) + f(v)$ and $f(v) \gets f(u) + f(v)$. 
\end{remark}

Additionally, we define a special class of row graphs.

\begin{definition}[Reversible Row Graph]
A row graph $(G\langle V, E\rangle,P,f)$ is called \textit{reversible} if and only if $P$ is invertible.
\end{definition}

Finally, we define another class of row graphs called basic row graph. For instance, a permutation matrix is represented by a basic reversible row graph.

\begin{definition}[Basic form]
A row graph $(G\langle V, E\rangle,P,f)$ is called \textit{basic} if and only if $size(f(u))=1$ for all $u \in V$.
\end{definition}

\begin{example}
\label{def: basic reversible row graph}
Example of a basic reversible row graph.
	\begin{figure}[H]
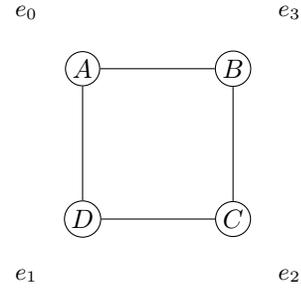

		\ctikzfig{basic_reversible_row_graph}
		\caption{Basic reversible row graph example} \label{fig: basic_reversible_row_graph}
	\end{figure}
\end{example}

\subsection{The Token Reduction Problem}
When using the row graph representation, we can treat each standard basis vector as a \textit{token}. Under this perspective, the task of decomposing a matrix in $\mathbb{F}_2$ can be visualized as reducing tokens on the row graph such that each node only has one token.

\begin{definition}[\textsc{Token Reduction}]
We define the \textsc{Token Reduction} problem being the decision problem of whether a row graph can be reduced to a basic row graph only by the adjacent node addition operations.
\end{definition}

\begin{theorem}
If a row graph is reversible, then it can be reduced to a basic graph.
\begin{proof}[Proof (Sketch)]
This is equivalent to the constrained CNOT synthesis problem. The Steiner Gauss method  can decompose any invertible $\mathbb{F}_2$ matrix given the architecture graph is connected. A row graph is a connected graph by Definition \ref{def: Row Graph}.

\end{proof}
\end{theorem}

\begin{definition}[\textsc{Opt Token Reduction}]
We also define the optimization version of the \textsc{Token Reduction} problem as \textsc{Opt Token Reduction}, which asks for the smallest sequence of row operations on adjacent nodes that transforms a given row graph to a basic row graph.
\end{definition}

\begin{theorem}
\textsc{Opt Token Reduction} is NP-hard for reversible row graphs.
\begin{proof}[Proof (Sketch)]
It is known that the \textsc{Swap Minimization Problem} is NP-hard \cite{inproceedings}. Token reduction for reversible row graph solves the \textsc{Swap Minimization Problem}.
\end{proof}
\end{theorem}

\begin{remark}
\label{remark:token reduction}
In this section, we proposed to reframe the constrained CNOT synthesis problem as token reduction on reversible row graphs. Although the two problems are equivalent, the token reduction representation provides us with a way to treat the constrained synthesis problem with more unity. In other words, instead of treating such problems as \textit{synthesis with graphical constraints}, we should treat them as \textit{synthesis on graphs}. 
\end{remark}

\section{Token Reduction Methods}
\label{sec: Token Reduction Methods}

We will spend the first two subsections to establish the theorems supporting the token reduction method and give an upper bound for the \textsc{Opt Token Reduction} problem for reversible row graphs. In Subsection \ref{sec: Heuristic Reduction Method}, we present an algorithm that outperforms the current testable benchmark. Finally, in Subsection \ref{sec: Token Reduction for General Circuits}, we discuss how this algorithm can be extended for general quantum circuits.

\subsection{Tree Reduction}
\label{sec: Tree Reduction}
The primitive operation on a row graph is the node addition operation. We want to develop a strategy that only uses node additions to reduce a row graph. In this section, we construct an abstraction of row additions called \textit{tree reduction}, which serves as the base operation for our task on hand. 

Firstly, we prove a critical lemma, which states that any node on a reversible row graph can be reduced to a single standard basis vector by a linear combination with other nodes under node addition.
 
\begin{lemma}
\label{lemma: reduction lemma}
For any $u \in V$ in a reversible row graph $(G\langle V, E\rangle,P,f)$ where $P$ is an $n\times n$ matrix, there exists a standard basis vector $y \in \mathbb{F}_2^n$ and a set of nodes $V' \subseteq V$ containing $u$, such that $\sum_{v \in V'}  f(v)  = y$.  
\begin{proof}
Suppose $V={v_1, v_2, \cdots , v_n}$, and $\textbf{r}_0, \cdots \textbf{r}_{n-1}$ are the row vectors in $P$. Now, to prove our lemma for the case $f(u) = \textbf{r}_k$ is equivalent to prove that there exists a standard basis vector $y$ such that the following equation has a solution with $x_k = 1$. Let's denote $\textbf{r}_{i,j}$ as the $j_{th}$ entry of $\textbf{r}_i$.

\begin{equation}
\label{eq: solution}
\begin{pmatrix}
\textbf{r}_{0,0}& \cdots & \textbf{r}_{k,0} & \cdots & \textbf{r}_{n-1,0}\\
\vdots & \cdots &\vdots & \cdots &\vdots \\
\textbf{r}_{0,n-1}& \cdots & \textbf{r}_{k,n-1} & \cdots & \textbf{r}_{n-1,n-1}\\
\end{pmatrix} 
\begin{pmatrix}x_0 \\ x_1 \\ \vdots\\  x_k \\ \vdots \\ x_{n-1} \end{pmatrix} = y
\end{equation}\\

Furthermore, equation (\ref{eq: solution}) with the constraint $x_k=1$ represents the following system.

\begin{equation}
\label{eq: system}
\begin{cases}
x_k &= 1\\
\textbf{r}_{0,0}x_0 + \cdots +  \textbf{r}_{k,0} x_k + \cdots + \textbf{r}_{n-1,0}x_{n-1}&= y_{0}\\
\vdots\\
\textbf{r}_{0,n-1}x_0 + \cdots +  \textbf{r}_{k,n-1} x_k + \cdots + \textbf{r}_{n-1,n-1}x_{n-1}&= y_{n-1}
\end{cases}
\end{equation}\\

Let's call all the expressions on the left side of the above equations except the first one $Z_i$ (e.g. $Z_0 = \textbf{r}_{0,0}x_0 + \cdots +  \textbf{r}_{k,0} x_k + \cdots + \textbf{r}_{n-1,0}x_{n-1}$). Notice that all $Z_i$s are linearly independent because the columns in $P$ are linearly independent. Therefore, $x_k$ must be a linear combination of all these $Z_i$s, let's call this combination set $Z'$. Since the system in (\ref{eq: system}) without $x_k =1$ always have a solution regardless of $y$, we can assign one of the expression in $Z'$ to 1 and all others to 0 hence we will have a solution for the entire system (\ref{eq: system}).

\end{proof}

\end{lemma}

\begin{definition}[Reduction tree]
Given a row graph $(G\langle V, E\rangle,P,f)$ with $|V|=n$, a root node $u \in V$, a standard basis vector $e \in \mathbb{F}_2^n$, and a set of nodes $V' \subseteq V$ containing $u$, such that $\sum_{v \in V'} f(u) = e$. A reduction tree is a Steiner Tree with root $u$ and terminals $V'$.
\end{definition}

\begin{definition}[Tree reduction]
Given a row graph $(G\langle V, E\rangle,P,f)$ and a reduction tree $T=(root, \mathcal{S}, \mathcal{K})$, where $\mathcal{S}$ is the set of terminals and $\mathcal{K}$ is the set of Steiner Points. A tree reduction is a procedure defined as Algorithm \ref{algo: tree reduction}.

\begin{algorithm}
\label{algo: tree reduction}
	\SetAlgoLined
	\KwIn{$T=(root, \mathcal{S}, \mathcal{K})$,$\mathcal{G}=(G\langle V, E\rangle,P,f)$ }
	\PostTraverse{$T$}{
		$u \gets$ current node\;
		\uIf{$u == root$}{
			\Break;		
		}
		\uElseIf{$u.parent \in \mathcal{K}$}{
			$swap(f(u),f(u.parent))$\;
			$\mathcal{K} \gets \mathcal{K} \cup \{u\}$\;
			$\mathcal{K} \gets \mathcal{K} \setminus u.parent$\;
		}
		\Else{
		$f(u.parent) \gets f(u.parent) + f(u) $
		}
	}
	 \caption{\textsc{Tree Reduction}}
\end{algorithm}

\end{definition}

Next, we give an example to show how tree reduction works.
\begin{example}
Consider the following reduction tree with assignment function $f$, where the root is $R$, the terminals are $\{R,B,D,E\}$, and the Steiner Points are the light blue nodes $\{A,C\}$.
	\begin{figure}[H]
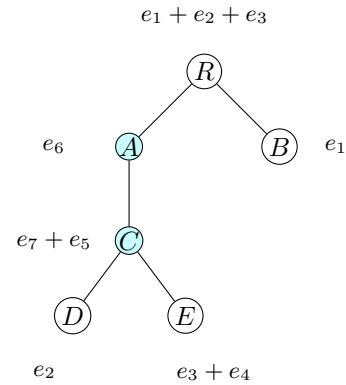

		\ctikzfig{tree_reduction_0}
		\caption{Reduction Tree} \label{fig: tree_reduction}
	\end{figure}

First notice that $f(R)+f(B)+F(D)+F(E) = e_4$. Next, we use the tree reduction procedure to reduce $f(R)$ to $e_4$.
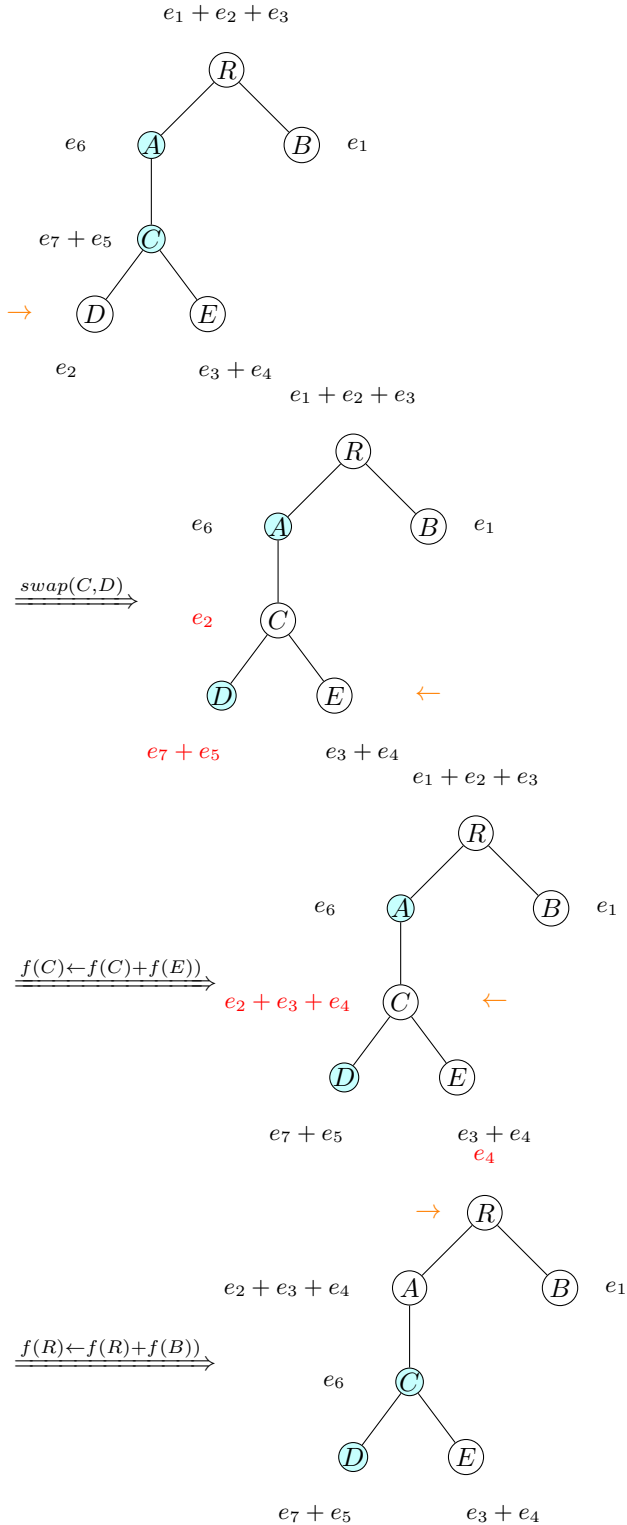
\begin{align*}
&\input{tree_reduction.tikz} \\
&\xRightarrow[]{swap(C,D)} \input{tree_reduction_2.tikz}\\
&\xRightarrow[]{f(C)\gets f(C)+f(E))} \input{tree_reduction_3.tikz} \\
%&\xRightarrow[]{swap(A,C)} \tikzfig{tree_reduction_4}\\
%&\xRightarrow[]{f(R)\gets f(R)+f(A))} \tikzfig{tree_reduction_5}\\
&\xRightarrow[]{f(R)\gets f(R)+f(B))} \input{tree_reduction_6.tikz}
\end{align*}
%
%$$\tikzfig{tree_reduction} \xRightarrow[]{swap(C,D)} \tikzfig{tree_reduction_2} $$
%$$\xRightarrow[]{f(C)\gets f(C)+f(E))} \tikzfig{tree_reduction_3} \xRightarrow[]{swap(A,C)} \tikzfig{tree_reduction_4}$$
%$$\xRightarrow[]{f(R)\gets f(R)+f(A))} \tikzfig{tree_reduction_5} \xRightarrow[]{f(R)\gets f(R)+f(B))} \tikzfig{tree_reduction_6} $$

%\vspace{12 pt}

The arrow indicates the currently visited node during the tree traversal.
\end{example}

%With Lemma \ref{lemma: reduction lemma}, we are now able to proof Theorem \ref{thm: reduction theorem}.
Next, we give a simple algorithm that reduce any reversible row graph with $O(n^2)$ node addition operations.
\begin{theorem}
\label{thm: reduction theorem}
Any reversible row graph with $|V|=n$ can be reduced to a basic row graph by $O(n^2)$ node addition operations.
\begin{proof}
We prove this theorem by constructing the Algorithm \ref{algo: simple reduction}. 

\begin{algorithm}
\label{algo: simple reduction}
	\SetAlgoLined
	\KwIn{$(G,P,f)$ }
	\For{$u\in V, f(u)>1$}{
		Find a standard basis vector $e \in \mathbb{F}_2^n$ and a set of nodes $V' \subseteq V$ containing $u$, such that $\sum_{v \in V'}  f(v)  = \{e\}$. We do this by solving equation (\ref{eq: solution}) for all $n$ standard basis vectors until we find a solution that contains $u$. Lemma \ref{lemma: reduction lemma} ensures such standard basis vector always exists\;
		Constructing a reduction tree $T$ with root as $u$ and terminals as $V'$\;
		Perform tree reduction on $T$\;
		Reversely perform all the reduction operations except the ones involve the root to make all non-root nodes back to their original vector assignments\;
	}
	 \caption{\textsc{Simple Token Reduction}}
\end{algorithm}

\begin{example}
\label{example: full recovery}
Making all non-root nodes back to their original vector assignments.
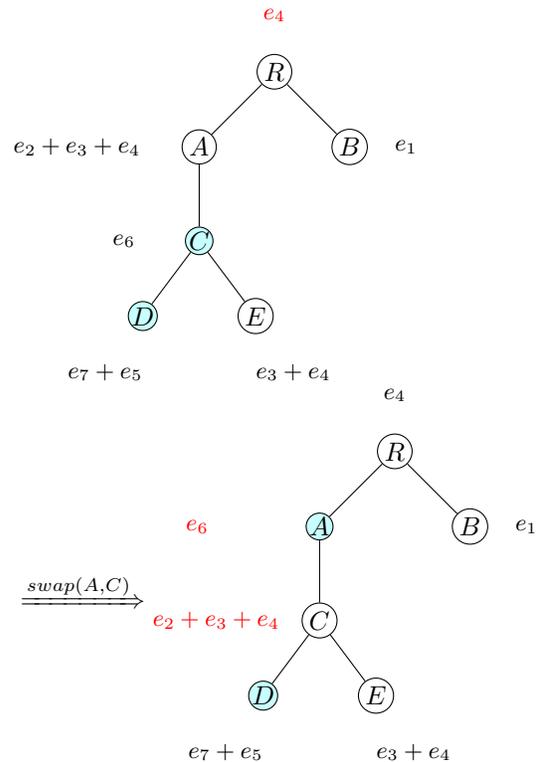
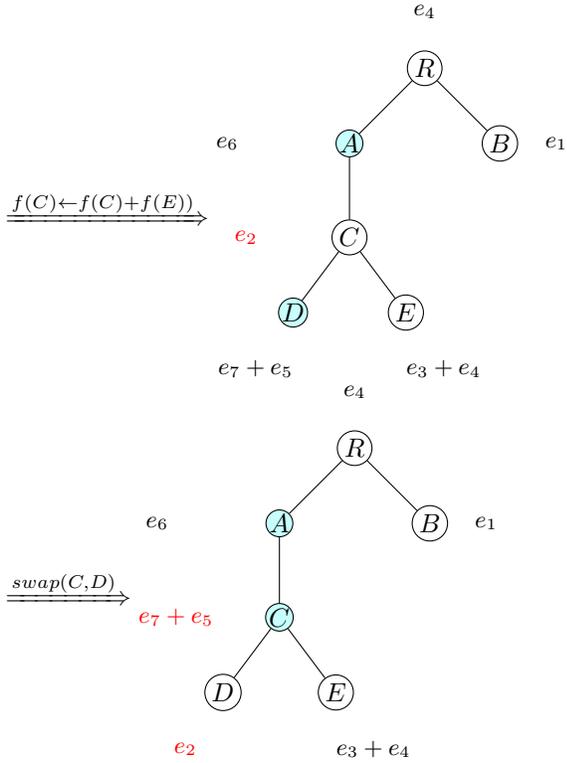
\begin{align*}
&\input{tree_reduction_r_0.tikz}\\
& \xRightarrow[]{swap(A,C)}\input{tree_reduction_r_1.tikz} \\
&\xRightarrow[]{f(C)\gets f(C)+f(E))}\input{tree_reduction_r_2.tikz}  \\
&\xRightarrow[]{swap(C,D)}  \input{tree_reduction_r_3.tikz} 
\end{align*}
%$$\tikzfig{tree_reduction_r_0} \xRightarrow[]{swap(A,C)}\tikzfig{tree_reduction_r_1} $$
%$$\xRightarrow[]{f(C)\gets f(C)+f(E))}\tikzfig{tree_reduction_r_2}  \xRightarrow[]{swap(C,D)}  \tikzfig{tree_reduction_r_3} $$
\end{example}

The original row graph can have none of its nodes being standard basis vector. For each of these $n$ nodes, we construct a Steiner tree which can have size of at most $n$. Then for each of these trees, there can be at most $n-2$ Steiner Points. Hence we need at most $3(n-2) + 1$ addition operations to reduce $u$. Finally the reversal step will need at most $3(n-2)$ addition operations. Hence the total operations required is at most $n(6(n-2)+1)$.
\end{proof}
\end{theorem}

The construction in the proof of Theorem \ref{thm: reduction theorem} provides us with a $O(n^2)$ upper bound for solutions in \textsc{Opt Token Reduction} for reversible row graphs. That is, Algorithm \ref{algo: simple reduction} synthesizes a $n$-qubit CNOT circuit with $O(n^2)$ CNOT gates. In the next section, we improve this algorithm by introducing a heuristic and using partial reversal rather than reverting all non-root nodes.

%
%\subsection{Naive Reduction Method}
%\label{sec: Naive Reduction Method}

\subsection{Heuristic Reduction Method}
\label{sec: Heuristic Reduction Method}

Before we describe the algorithm, we define two heuristic functions.

\begin{definition}[Cost function]
Given any node $v$ in a reversible row graph $(G\langle V, E\rangle,P,f)$, and a standard basis vector $e$ in $\mathbb{F}_2^n$. If $v$ can be reduced to $e$ by a linear combination with other nodes (see Lemma \ref{lemma: reduction lemma}), then let $T$ be the reduction tree that reduces the node $v$ to $e$. $cost(v,e)$ is defined as the number of node addition operations to reduce $v$ to $e$ on $T$, plus the number of operations required to recover any standard basis vector that were transformed to non-basis vector during the reduction.
If $v$ cannot be reduced to $e$ by a linear combination with other nodes, then $cost(v,e) = \infty$.
\end{definition}

To exactly compute the cost function, we define the following two algorithms. Algorithm \ref{algo: tree reduction with tracking} is identical to Algorithm \ref{algo: tree reduction} but with additional information indicating which nodes need to be recovered.

\vspace{12 pt}
\begin{algorithm}
\label{algo: tree reduction with tracking}
	\SetAlgoLined
	\KwIn{$T=(root, \mathcal{S}, \mathcal{K}), \mathcal{G}=(G\langle V, E\rangle,P,f)$ }
	$R \gets \{\}$\;
	$operations \gets []$\;
	\PostTraverse{$T$}{
		$u \gets$ current node\;
		\uIf{$u == root$}{
			\Break;		
		}
		\uElseIf{$u.parent \in \mathcal{K}$}{
			\If{$u \in R$}{
				$R \cup \{u.parent\}$\;
				$R \setminus \{u\}$\;
			}
			$swap(f(u),f(u.parent))$\;
			$\mathcal{K} \gets \mathcal{K} \cup \{u\}$\;
			$\mathcal{K} \gets \mathcal{K} \setminus u.parent$\;
			$operations.append(``SWAP", u, u.parent)$\;
		}
		\Else{
			\If{$size(f(u.parent)) == 1$}{
				$R \cup \{u.parent\}$\;
			}
		$f(u.parent) \gets f(u.parent) + f(u) $\;
		$operations.append(``ADD", u.parent, u)$\;
		}
	}
	\Return $operations$, $R$\;
	 \caption{\textsc{Tree Reduction with Tracking}}
\end{algorithm}

\vspace{12 pt}
\begin{algorithm}
\label{algo: partial recovery}
	\SetAlgoLined
	\KwIn{$operations, R, T=(root, \mathcal{S}, \mathcal{K}), \mathcal{G}=(G\langle V, E\rangle,P,f)$ }
	$recover\_ops \gets []$\;
	\For{$opType, node1, node2 \in reversed(operations)$}{
		\uIf{$opType == ``ADD"$ and $node1 \in R$}{
			$f(node1) \gets f(node1) + f(node2) $\;
			$recover\_ops.append(``ADD", node1, node2)$\;
			\If{$size(f(node1))==1$}{
				$R \setminus \{node1\}$\;
			}
		}
		\uElseIf{$opType == ``SWAP"$ and $node2 \in R$}{
			$swap(f(node1),f(node2))$\;
			$R \cup \{node1\}$\;
			$R \setminus \{node2\}$\;
			$recover\_ops.append(``SWAP", node1, node2)$\;
			
		}
	}
	\Return $recover\_ops$

	 \caption{\textsc{Reduction Recovery}}
\end{algorithm}
\vspace{12 pt}

Algorithm \ref{algo: partial recovery} is similar to the operation reversal procedure described in Example \ref{example: full recovery}, but only reversely performs operations that are necessary to recover those standard basis vectors that were transformed to non-basis vector during the reduction.

Consequently, the cost function can be computed directly by performing Algorithm \ref{algo: tree reduction with tracking} and Algorithm \ref{algo: partial recovery}, then counting the total number of addition operations (a SWAP counts for 3 addition operations).
\begin{definition}[Loss function]
\label{def: loss function}
For a reversible row graph $\mathcal{G}=(G\langle V, E\rangle,P,f)$, we compute the loss function by the following steps:

\begin{enumerate}
\item Let $B$ be the set of standard basis vectors in $\mathbb{F}_2^n$. For all $(u,e) \in V \times B$ construct the following table:
\begin{table}[H]
\centering
\begin{tabular}{|
>{\columncolor[HTML]{E4E0E0}}c |c|c|c|c|c|}
\hline
         & \cellcolor[HTML]{E4E0E0}$e_1$ & \cellcolor[HTML]{E4E0E0}$e_2$ & \cellcolor[HTML]{E4E0E0}$e_3$ & \cellcolor[HTML]{E4E0E0}$\cdots$ & \cellcolor[HTML]{E4E0E0}$e_n$ \\ \hline
$u_1$    & $cost(u_1, e_1)$              & $cost(u_1, e_2)$              & $cost(u_1, e_3)$              & $\cdots$                         & $cost(u_1, e_n)$              \\ \hline
$u_2$    & $cost(u_2, e_1)$              & $cost(u_2, e_2)$              & $cost(u_2, e_3)$              & $\cdots$                         & $cost(u_2, e_n)$              \\ \hline
$u_3$    & $cost(u_3, e_1)$              & $cost(u_3, e_2)$              & $cost(u_3, e_3)$              & $\cdots$                         & $cost(u_3, e_n)$              \\ \hline
$\vdots$ & $\vdots$                      & $\vdots$                      & $\vdots$                      & $\vdots$                         & $\vdots$                      \\ \hline
$u_n$    & $cost(u_n, e_1)$              & $cost(u_n, e_2)$              & $cost(u_n, e_3)$              & $\cdots$                         & $cost(u_n, e_n)$              \\ \hline
\end{tabular}
\caption{Cost table}
\label{table: cost table}
\end{table}

\item Let $h$ be an bijective assignment function $h: V \to B$, such that minimizes $\sum_{u\in V} cost(u,h(u))$. And the loss function is defined as $$Loss(\mathcal{G}) = \sum_{u\in V} cost(u,h(u))$$

\item The function $h$ is solved by using the Hungarian algorithm \footnote{This project uses an open source implementation available at \href{http://software.clapper.org/munkres/}{http://software.clapper.org/munkres/} }. 
\end{enumerate}
\end{definition}

\begin{theorem}
Given a reversible row graph $\mathcal{G}=(G\langle V, E\rangle,P,f)$. $cost(u,h(u))$ is finite for all $u \in V$, where $h$ is defined in Definition \ref{def: loss function}.
\begin{proof}
This is equivalent to prove a perfect matching exists in a bipartite graph. Hall's marriage theorem states that for a bipartite graph with vertex set $V_1$ and vertex set $V_2$, a perfect matching exists if and only if $|neighbours(S)| \geq |S| \text{ for all } S \subset V_1 $ (Hall's condition). That is, we want to prove for every set $S \subset B$, $|neighbours(S)| \geq |S|$. We prove it by contradiction. Each standard basis vector in $B$ is a linear combination of some the row vectors on some nodes. Suppose $|neighbours(S)| < |S|$ for some set $S \subset B$, this implies the system has more equations than variables (i.e. equations are of the form $e_i=f(u_1)+f(u_2)+\cdots + f(u_k)$ for some $u_j \in V$), hence there exists a linear combination of row vectors that gives two different results, which is clearly false.
\end{proof}
\end{theorem}

Finally, we show a heuristic algorithm that outperforms the current benchmark when combined with a simple post-processing procedure.

\vspace{12 pt}
\begin{algorithm}
	\label{algo: heuristic token reduction}
	\SetAlgoLined
	\KwIn{$\mathcal{G}=\langle G,P,f \rangle$ }
	\While{not in basic form}{
		We compute the cost table $\mathcal{T}$ \ref{table: cost table}\;
		Let $S$ be the list of $(u,e)$ pairs in $\mathcal{T}$ with the smallest cost\;
		$(u_{min}, e_{min}) \gets null$\;
		$loss_{min} \gets \infty$\;
		\For{$(u,e) \in S$}{
			Reduce $u$ to $e$ using Algorithm \ref{algo: tree reduction with tracking} and Algorithm \ref{algo: partial recovery}\;
			\If{$Loss(\mathcal{G}) < loss_{min}$}{
				$loss_{min} \gets Loss(\mathcal{G})$\;
				$(u_{min}, e_{min}) \gets (u,e)$\;
			}
			Reverse \textit{all} operations\;
		}
		Reduce $u_{min}$ to $e_{min}$ by Algorithm \ref{algo: tree reduction with tracking}\;
		Recover standard basis vectors by Algorithm \ref{algo: partial recovery}\; 
	}
	 \caption{\textsc{Heuristics Token Reduction}}
\end{algorithm}
\vspace{12 pt}

The routed circuit is then subjected to the following post-processing procedure:
\begin{enumerate}
\item SWAP synthesis: for each SWAP gate, convert it to CNOT gates in an orientation (see Equation (\ref{eq: swap to cnot})) that can hopefully cancel some adjacent CNOT gates.
\item CNOT cancellation: for each CNOT gate, try to cancel it with other CNOT gates by Equation (\ref{eq: cnot identity}).\footnote{The procedure used in this project is the basic\_optimization method implemented in \textit{pyzx} \cite{kissinger2019pyzx}}
\end{enumerate} 

Firstly, notice that we select $u_{min}, e_{min}$ that have the smallest cost; this ensures that the algorithm reduces exactly one node for each iteration. This also helps to limit the number of times to compute the loss function. Secondly, minimizing the loss function helps the algorithm select a reduction that transforms the token graph such that the remaining iterations require few operations.

\subsection{Complexity Analysis}
\label{sec: Complexity Analysis}

In terms of the final circuits' gate-count complexity, Algorithm \ref{algo: heuristic token reduction} reduces one non-standard basis vector for each iteration. The reduction and recovery procedure cost $O(n)$ node addition operations ($n$ is the number of qubits). Therefore, the complexity is $O(n^2)$. 

Next, we analyze the computational complexity. First, notice that the approximation Steiner Tree algorithm we used has a complexity of $O(n^3)$ (see Appendix \ref{app:steiner}). Then for the cost table, we compute one Steiner tree for each standard basis vector $e$. Then we compute the cost for each $(u,e)$ pair by selecting different nodes as the root. Hence the cost for constructing a cost table is of $O(n^4)$.

Inside the while-loop, the computational bottleneck lies in the for-loop started at line $7$. The set $S$ can at most have $n^2$ elements; for each $(u,e)$ pair, we have to compute a loss after performing the reduction. The complexity of computing the loss function is dominated by the cost of constructing a cost table. Hence each iteration in the for-loop has a complexity of $O(n^4)$, and the complexity for the entire for-loop is $O(n^6)$.

Finally, in the worst case, we need to reduce $n$ nodes in total. Hence the total computational complexity for Algorithm \ref{algo: heuristic token reduction} is $O(n^7)$.

\subsection{Token Reduction for General Circuits}
\label{sec: Token Reduction for General Circuits}
Recall that our goals for designing a CNOT synthesis method are: firstly, synthesize using as few operations as possible; and secondly, synthesize to an output mapping that benefits the rest of the circuit. In the previous section, we have addressed the first goal by introducing the heuristic token reduction algorithm. Here we discuss a possible direction to address the second goal. 

One characteristic of the heuristic token reduction method is that we have some freedom for selecting a reduction tree in each iteration. For general circuits, the candidate evaluation process can also be adjusted to take into account the difficulty of synthesizing the next CNOT block in the circuit.  We already have an approximation for measuring the difficulty of synthesizing a block of CNOT gates, the loss function defined in Definition \ref{def: loss function}. Therefore we only need a way to construct the next CNOT block as a row graph in order to compute the loss function. Constructing a row graph requires only an initial mapping. We can do this by guessing an output permutation that a reduction candidate would most likely lead to. This guessing can be facilitated by the cost table and the assignment function $h$ defined in Definition \ref{def: loss function}.

\section{Result}
\subsection{Benchmark}
\label{sec: Benchmark}

A recent study demonstrated a constrained CNOT synthesis approach that relies on solving \textit{the syndrome decoding problem} \cite{10.1007/978-3-030-52482-1_11}. Their test results showed improvement against the Steiner Gauss method on 15 different architectures; hence their method can be considered the current state-of-the-art. However, a valid comparison with their method is not possible because their code is not accessible, and their test circuits are not published. Judging only by the available data shown in their paper, the token reduction method performs better for four of the the five architectures tested in this project. Additionally, their method only works for architectures containing a Hamiltonian Path, whereas our approach works with all connected architectures.

As a result, we chose the Steiner Gauss method\cite{kissinger2019cnot} as the benchmark in order to have a direct comparison. Their method has been considered the state-of-the-art until the work mentioned above was published; and more importantly, their code is publicly accessible \footnote{Their source code is available at the Github repository: \href{https://github.com/Quantomatic/pyzx}{pyzx}}.

\subsection{Comparison Method}
\label{sec: Comparison Method}

For each of the five architectures tested, we consider CNOT circuits with either 4,8,16,32,64,128 or 256 gate counts. Then, for each of these gate counts, we generated 100 random CNOT circuits to perform synthesis and compared the results between the Steiner Gauss method and the token reduction method. For each random circuit, we used the same arbitrary initial mapping for both methods (See Appendix \ref{app:Example Architectures}). Moreover, after performing the token reduction method for each circuit, we verified the the output circuit was equivalent (up to a permutation) to the original circuit by direct matrix comparison. Our code is entirely written in Python; even though the implementation hasn't been well optimized, the token reduction program still took less than 20 minutes to test all 3500 circuits on a consumer-grade laptop.

Table \ref{tab: method comparison} shows a comparison between the Steiner Gauss method and the token reduction method for 5 different architectures. The first column shows the architectures tested in this comparison. The second column titled ``\#" shows the gate counts in the tested circuits. The third and fourth columns show the average of the output CNOT counts for the Steiner Gauss and the token reduction methods respectively. The fifth, sixth, and seventh columns show the average, minimum and maximum percent improvements in total CNOT count of our approach vs. the Steiner Gauss method. The last column shows the percentage of the tested circuits in which our approach outperformed the Steiner Gauss method. All of the 5 architecture graphs are listed in Appendix \ref{app:Example Architectures}.

\begin{table*}[htbp]
  \small
  \centering
  \caption{Methods comparison}
    \begin{tabular}{|c|r|r|r|r|r|r|r|}
    \hline
          &       & \multicolumn{2}{p{6.33em}|}{Average output CNOT count} & \multicolumn{4}{p{13.495em}|}{Saving: TR against Steiner} \bigstrut\\
    \hline
    \rowcolor[rgb]{ .816,  .808,  .808} \multicolumn{1}{|l|}{Architecture} & \multicolumn{1}{l|}{\#} & \multicolumn{1}{l|}{Steiner} & \multicolumn{1}{l|}{TR} & \multicolumn{1}{l|}{Mean} & \multicolumn{1}{l|}{Max} & \multicolumn{1}{l|}{Min} & \multicolumn{1}{l|}{Positive} \bigstrut\\
    \hline
    \multirow{7}[14]{*}{9-square} & 4     & 15.51 & 11.67 & 21.56\% & 66.67\% & -40.00\% & 93.00\% \bigstrut\\
\cline{2-8}          & 8     & 29.7  & 20.35 & 28.59\% & 64.44\% & -50.00\% & 93.00\% \bigstrut\\
\cline{2-8}          & 16    & 44.78 & 31.08 & 29.09\% & 54.76\% & -9.38\% & 95.00\% \bigstrut\\
\cline{2-8}          & 32    & 55.84 & 39.31 & 28.94\% & 51.35\% & -8.11\% & 99.00\% \bigstrut\\
\cline{2-8}          & 64    & 60.43 & 42    & 29.74\% & 56.79\% & 2.04\% & 100.00\% \bigstrut\\
\cline{2-8}          & 128   & 60.38 & 41.33 & 30.91\% & 54.93\% & 0.00\% & 100.00\% \bigstrut\\
\cline{2-8}          & 256   & 59.67 & 43.16 & 27.14\% & 45.95\% & -2.22\% & 99.00\% \bigstrut\\
    \hline
    \multirow{7}[14]{*}{16-square} & 4     & 28.24 & 21.29 & 20.88\% & 63.41\% & -32.43\% & 90.00\% \bigstrut\\
\cline{2-8}          & 8     & 56.93 & 40.11 & 27.40\% & 58.06\% & -15.09\% & 97.00\% \bigstrut\\
\cline{2-8}          & 16    & 98.95 & 64.08 & 33.46\% & 64.78\% & -11.67\% & 98.00\% \bigstrut\\
\cline{2-8}          & 32    & 154.93 & 109.69 & 28.36\% & 48.26\% & 6.19\% & 100.00\% \bigstrut\\
\cline{2-8}          & 64    & 196.73 & 149.54 & 23.85\% & 42.20\% & 8.65\% & 100.00\% \bigstrut\\
\cline{2-8}          & 128   & 205.41 & 165.08 & 19.35\% & 34.51\% & -3.47\% & 99.00\% \bigstrut\\
\cline{2-8}          & 256   & 204.35 & 163.48 & 19.78\% & 37.13\% & 5.76\% & 100.00\% \bigstrut\\
    \hline
    \multirow{7}[14]{*}{ibmqx5} & 4     & 39.84 & 30.83 & 20.14\% & 53.19\% & -25.00\% & 88.00\% \bigstrut\\
\cline{2-8}          & 8     & 71.47 & 52.47 & 24.18\% & 61.11\% & -76.47\% & 94.00\% \bigstrut\\
\cline{2-8}          & 16    & 126.03 & 88.13 & 28.29\% & 58.79\% & -50.00\% & 98.00\% \bigstrut\\
\cline{2-8}          & 32    & 183.89 & 136.95 & 24.60\% & 50.24\% & -7.95\% & 99.00\% \bigstrut\\
\cline{2-8}          & 64    & 230.71 & 186.87 & 18.65\% & 39.07\% & -4.69\% & 97.00\% \bigstrut\\
\cline{2-8}          & 128   & 242.24 & 199.79 & 17.28\% & 32.95\% & 2.75\% & 100.00\% \bigstrut\\
\cline{2-8}          & 256   & 245.64 & 201.48 & 17.83\% & 31.01\% & 2.18\% & 100.00\% \bigstrut\\
    \hline
    \multirow{7}[14]{*}{rigetti-16q-aspen} & 4     & 59.37 & 30.32 & 43.60\% & 81.44\% & -24.00\% & 98.00\% \bigstrut\\
\cline{2-8}          & 8     & 101.24 & 56.5  & 41.72\% & 66.22\% & -22.45\% & 98.00\% \bigstrut\\
\cline{2-8}          & 16    & 166.03 & 96.81 & 40.61\% & 60.11\% & -17.53\% & 99.00\% \bigstrut\\
\cline{2-8}          & 32    & 223.05 & 156.91 & 29.26\% & 49.38\% & 9.90\% & 100.00\% \bigstrut\\
\cline{2-8}          & 64    & 260.43 & 209.42 & 19.34\% & 34.73\% & -5.38\% & 99.00\% \bigstrut\\
\cline{2-8}          & 128   & 271.12 & 226.81 & 16.15\% & 31.09\% & 1.19\% & 100.00\% \bigstrut\\
\cline{2-8}          & 256   & 271.5 & 228.57 & 15.62\% & 29.24\% & -4.47\% & 98.00\% \bigstrut\\
    \hline
    \multirow{7}[14]{*}{ibm-q20-tokyo} & 4     & 23.75 & 17.72 & 22.87\% & 52.17\% & -16.67\% & 97.00\% \bigstrut\\
\cline{2-8}          & 8     & 50.89 & 33.25 & 32.11\% & 62.50\% & -12.00\% & 96.00\% \bigstrut\\
\cline{2-8}          & 16    & 99.57 & 64.88 & 31.93\% & 59.15\% & -3.17\% & 97.00\% \bigstrut\\
\cline{2-8}          & 32    & 177.28 & 116.83 & 33.24\% & 53.44\% & 3.33\% & 100.00\% \bigstrut\\
\cline{2-8}          & 64    & 254.34 & 191.83 & 24.24\% & 48.44\% & 3.90\% & 100.00\% \bigstrut\\
\cline{2-8}          & 128   & 290.99 & 233.66 & 19.54\% & 33.54\% & 0.00\% & 100.00\% \bigstrut\\
\cline{2-8}          & 256   & 293.34 & 235.83 & 19.46\% & 33.87\% & 6.94\% & 100.00\% \bigstrut\\
    \hline
    \end{tabular}
  \label{tab: method comparison}
\end{table*}

\section{Discussion}

\subsection{Summary}
\label{sec: Summary}

In this study, we introduced a CNOT synthesis method for quantum computers with restricted connectivity. Our method utilizes the observation that a qubit routing procedure doesn’t require the routed circuit to have the same permutation for its output as for its input. As long as the permutation for the output is known, we can re-order the output after the quantum computation finishes. Our approach synthesizes a CNOT circuit to a permutation that is decided on the fly in order to find the best output permutation that is the easiest for the synthesis procedure. This approach consistently outperforms the benchmark method \cite{kissinger2019cnot}, which always synthesizes a CNOT circuit such that its input and output have the same permutation. Moreover, our approach works for all connected quantum computer architectures without any graphical assumptions.

Another important characteristic of our approach is that we have some control over the output’s permutation of a routed circuit. This property could prove beneficial when we extend our method for general quantum circuits. A general quantum circuit can be converted and partitioned into blocks of CNOT gates and blocks of one-qubit gates. Therefore, the output’s permutation for one CNOT sub-circuit will have a substantial effect on the difficulty to synthesize the remaining CNOT sub-circuits. 

Furthermore, we also proposed a new representation for working with the constrained CNOT synthesis problem. This representation, which we call the row graph, is more visually appealing than the previously used matrix \& graph representation. We expect more synthesis techniques to be inspired by this row graph representation. 

\subsection{Future Work}
\label{sec: Future Work}

In Section \ref{sec: Qubit Routing Problem} we mentioned that there a second issue associated with restricted connectivity for NISQ devices. This issue states that a CNOT gate can only be applied to adjacent qubits with certain directions. We briefly explained that this issue could be addressed by converting a CNOT with one direction to a CNOT with the opposite direction using Hadamard gates (Equation \ref{eq: cnot direction}). However, future studies might want to take this issue into account while designing the synthesis algorithm. For example, when selecting a tree reduction, we might want to prefer a tree whose node addition operations mostly follow the architecture's direction constraints. 

In addition to designing a look-ahead mechanism to optimize the output's permutation (see Section \ref{sec: Token Reduction for General Circuits}), we also need to find a suitable method for optimizing the initial mapping. The method in \cite{kissinger2019cnot}  uses a genetic algorithm to find an initial mapping. For each selection iteration, they run the entire synthesis algorithm for all populations to compute the fitness scores. However, our method has a larger run-time complexity; hence running the entire synthesis procedure to compute fitness score would require significantly more time. Therefore, we might want to design a fitness function that is easier to compute while accurately reflect the score for each mapping in the population.

Lastly, the approximate minimum Steiner Tree algorithm (see Appendix \ref{app:steiner}) used in this project is not only inefficient but also not tailor-made for our tree reduction procedure. Two examples of this issue can be seen in Appendix \ref{app:Examples for inefficient Steiner Trees}. Therefore, designing a better algorithm to construct Steiner trees would have a straightforward improvement to our algorithm. We expect that our algorithm will become more affected by a non-ideal Steiner Tree algorithm for larger architectures. Even though finding a minimum Steiner Tree is an NP-complete problem on general graphs, we can focus our efforts on designing a heuristic algorithm for each specific architecture.

\section*{Acknowledgment}

This work is conducted as the the author's master's thesis. We thank Professor Kissinger for his valuable inputs and patient guidance.

\bibliography{refs.bib}{}
\bibliographystyle{IEEEtran}
%\begin{thebibliography}{00}
%\bibitem{b1} G. Eason, B. Noble, and I. N. Sneddon, ``On certain integrals of Lipschitz-Hankel type involving products of Bessel functions,'' Phil. Trans. Roy. Soc. London, vol. A247, pp. 529--551, April 1955.
%\bibitem{b2} J. Clerk Maxwell, A Treatise on Electricity and Magnetism, 3rd ed., vol. 2. Oxford: Clarendon, 1892, pp.68--73.
%\bibitem{b3} I. S. Jacobs and C. P. Bean, ``Fine particles, thin films and exchange anisotropy,'' in Magnetism, vol. III, G. T. Rado and H. Suhl, Eds. New York: Academic, 1963, pp. 271--350.
%\bibitem{b4} K. Elissa, ``Title of paper if known,'' unpublished.
%\bibitem{b5} R. Nicole, ``Title of paper with only first word capitalized,'' J. Name Stand. Abbrev., in press.
%\bibitem{b6} Y. Yorozu, M. Hirano, K. Oka, and Y. Tagawa, ``Electron spectroscopy studies on magneto-optical media and plastic substrate interface,'' IEEE Transl. J. Magn. Japan, vol. 2, pp. 740--741, August 1987 [Digests 9th Annual Conf. Magnetics Japan, p. 301, 1982].
%\bibitem{b7} M. Young, The Technical Writer's Handbook. Mill Valley, CA: University Science, 1989.
%\end{thebibliography}

\appendices
\section{Heuristic Algorithm for Steiner Trees}
\label{app:steiner}

Assume the input graph $G$ has its vertices labeled by $[0,1,\cdots,n-1]$.

\begin{algorithm}
\SetAlgoLined
\KwIn{A architecture graph $G\langle V, E \rangle$, set of vertices $S$, root}
\KwOut{A Steiner tree $T$}
	\# For each edge $(u,v)$ add an edge $(v,u)$ \;
	$newEdges = \{\}$\;
	\ForEach{ $(u,v) \in E_{G}$ }{
		$newEdges.add((v,u))$\;	
	}
	$E_{G}.union(newEdges)$\;
	$Dist,Successors$=\textsc{Floyd-WarshallWithPath}($G'$, $(u,v) \mapsto 1$)\;
	
	\# Find a pair of vertices in $S$ with minimal distance\;

	$(u,v) = \textsc{NearestNeighbours}(S,S,Dist)$;\
	
	\# The heuristic\;
	
	$S2 = S.copy()$\;
	\# Tree like graph\;
	$T = graph()$\;
	$path$ = \textsc{PathFromSuccessors}(Successors,u, v)\;
	\# Add the path to the tree and remove vertices from $S$ \;
	\textsc{AddPathToGraph}(T,path)\;
	$S2 = S2 \setminus  V_T$ \;	
	\While{ $S2.isNotEmpty()$}{
		$(u,v) = \textsc{Nearest Neighbours}(S2,V_{T},Dist)$;\
		$path$ = \textsc{PathFromSuccessors}(Successors,u, v)\;
		\textsc{AddPathToGraph}(T,path)\;
		$S2 = S2 \setminus  V_T$ \;	
	}
	\# Converted the tree like graph to a rooted tree by BFS\;
	RT = \textsc{Treefy}(T,root)\;
		
\Return{RT}
 \caption{\textsc{genSteiner}}
\end{algorithm}

\newpage

\begin{algorithm}
\SetAlgoLined
\KwIn{A directed weighted graph $G\langle V,E  \rangle$ where $V=\{0,1,2,\cdots,|V-1|\}$, a function $weight()$}
\KwOut{A $|V| \times |V|$ distance matrix $D$, a $|V| \times |V|$ Successor matrix $S$}
 Construct a $|V| \times |V|$ distance matrix $D$ with 0 for every diagonal entry, and other entries has $\infty$\;
 Construct a $|V| \times |V|$ Successor matrix $S$ with i (e.g $i_{th}$ row) for every diagonal entry, every other entry initialized null\;
 \ForEach{$(u,v) \in E$}{
 	$D[u][v] = weight((u,v))$
 	$S[u][v] = v$
 }
 \For{$k=0;k<|V|;k=k+1$}{
 	\For{$i=0;i<|V|;i=i+1$}{
 		\For{$j=0;j<|V|;j=j+1$}{
 			\If{$D[i][j] > D[i][k] + D[k][j]$}{
 				$D[i][j] = D[i][k] + D[k][j]$\;
 				$S[i][j] = S[i][k]$
 			}
 		}
 	}
 }
\Return{D,S}
 \caption{\textsc{Floyd-WarshallWithPath}}
\end{algorithm}

\vspace{12 pt}

\begin{algorithm}
\SetAlgoLined
\KwIn{Successor matrix $S$,  vertex $u$, vertex $v$}
\KwOut{A path list object $path$ contains the path from $u$ to $v$}
	\If{$S[u][v] == null$}{
		\Return{$[]$}	
	}
	$path = [u]$\;
	$x = u$\;
	\While{$x \neq v$}{
		$x=S[x][v]$	\;
		$path.append(x)$\;
	}
	
\Return{$path$}
 \caption{\textsc{PathFromSuccessors}}
\end{algorithm}

\section{Examples of inefficient Steiner Trees}
\label{app:Examples for inefficient Steiner Trees}
\begin{example}
Given root $Q2$, terminals $\{Q2, Q6, Q9, Q11\}$, the following tree requires $9$ node addition operations to reduce.
	\begin{figure}[H]
		\ctikzfig{inefficient_tree}
		\caption{Inefficient Tree (1) on 16-square architecture} \label{fig: inefficient_tree}
	\end{figure}
while an efficient tree would be:
	\begin{figure}[H]
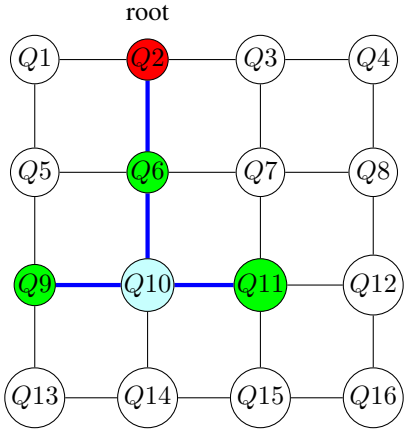

		\ctikzfig{efficient_tree}
		\caption{Efficient Tree (1) on 16-square architecture} \label{fig: efficient_tree}
	\end{figure}
The efficient tree requires $6$ node addition operations to reduce; hence a $33.33\%$ improvement.
\end{example}

\begin{example}
The following tree reduces $Q2$ to $e_8$ with $7$ node addition operations. However, except $Q1$ and the root, all other terminal nodes will then have non-unit vectors after the reduction. Hence we have $6$ nodes to be recovered. 
	\begin{figure}[H]
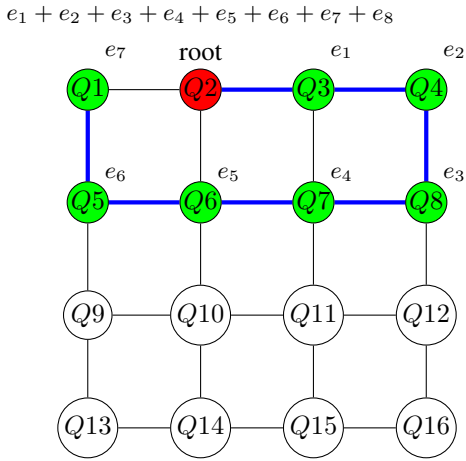

		\ctikzfig{inefficient_tree_2}
		\caption{Inefficient Tree (2) on 16-square architecture} \label{fig: inefficient_tree_2}
	\end{figure}

The second tree also requires $7$ node addition operations to reduce. However, we would only have to recover $Q1, Q3$ and $Q4$ after the reduction.
	\begin{figure}[H]
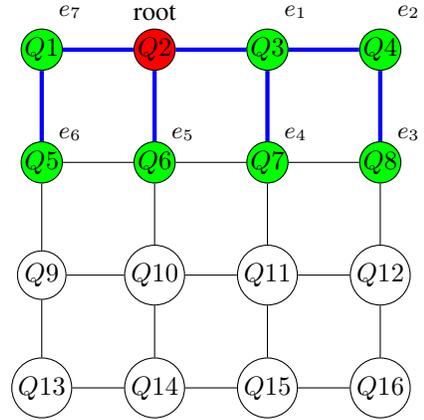

		\ctikzfig{efficient_tree_2}
		\caption{Efficient Tree (2) on 16-square architecture} \label{fig: efficient_tree_2}
	\end{figure}
\end{example}

\section{Example Architectures}
\label{app:Example Architectures}

In this section we provides the architecture graphs for the 5 architectures tested in Chapter \ref{sec: Comparison Method}. Notice that we ignore the directions of the edges in some architectures for the reason explained in Section \ref{sec: Qubit Routing Problem}.

We also include the initial mappings for each architecture used in the comparison test. For each $n$-qubit circuit, we assume the wires are labeled as $w_1, w_2 , \cdots , w_n$ from the top to the bottom.

\begin{itemize}

\item 9-square:\\
 	\begin{figure}[H]
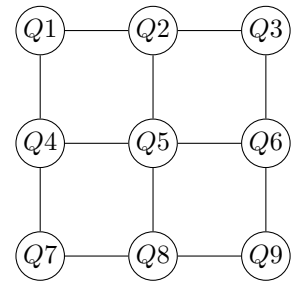

		\ctikzfig{9grid}
		\caption{9-square} \label{fig: 9grid}
	\end{figure}

$$M_0=\begin{cases}
w_1 \mapsto Q1, & w_2 \mapsto Q2 \\
w_3 \mapsto Q3 , & w_4 \mapsto Q6 \\
w_5 \mapsto Q5 , & w_6 \mapsto Q4 \\
w_7 \mapsto Q7 , & w_8 \mapsto Q8 \\
w_9 \mapsto Q9 \\
\end{cases}$$

\item 16-square:\\
 	\begin{figure}[H]
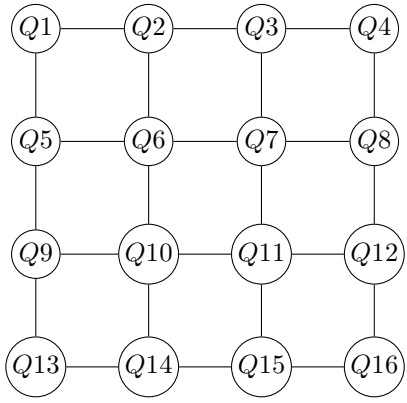

		\ctikzfig{16grid}
		\caption{16-square} \label{fig: 16grid}
	\end{figure}
	
$$M_0=\begin{cases}
w_1 \mapsto Q1, & w_2 \mapsto Q2 \\
w_3 \mapsto Q3, & w_4 \mapsto Q4 \\
w_5 \mapsto Q8 , & w_6 \mapsto Q7 \\
w_7 \mapsto Q6 , & w_8 \mapsto Q5 \\
w_9 \mapsto Q9 , & w_{10} \mapsto Q10 \\
w_{11} \mapsto Q11 , & w_{12} \mapsto Q12 \\
w_{13} \mapsto Q16 , & w_{14} \mapsto Q15 \\
w_{15} \mapsto Q14 , & w_{16} \mapsto Q13 \\
\end{cases}$$

\item rigetti-16q-aspen:\\
 	\begin{figure}[H]
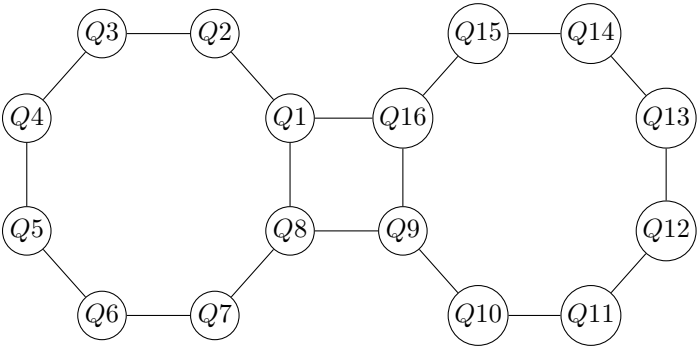

		\ctikzfig{rigetti_16q_aspen}
		\caption{rigetti-16q-aspen} \label{fig: rigetti_16q_aspen}
	\end{figure}
	
$$M_0=\begin{cases}
w_1 \mapsto Q1, & w_2 \mapsto Q2 \\
w_3 \mapsto Q3, & w_4 \mapsto Q4 \\
w_5 \mapsto Q5 , & w_6 \mapsto Q6 \\
w_7 \mapsto Q7 , & w_8 \mapsto Q8 \\
w_9 \mapsto Q9 , & w_{10} \mapsto Q10 \\
w_{11} \mapsto Q11 , & w_{12} \mapsto Q12 \\
w_{13} \mapsto Q13 , & w_{14} \mapsto Q14 \\
w_{15} \mapsto Q15 , & w_{16} \mapsto Q16 \\
\end{cases}$$

\item ibm-q20-tokyo:\\
 	\begin{figure}[H]
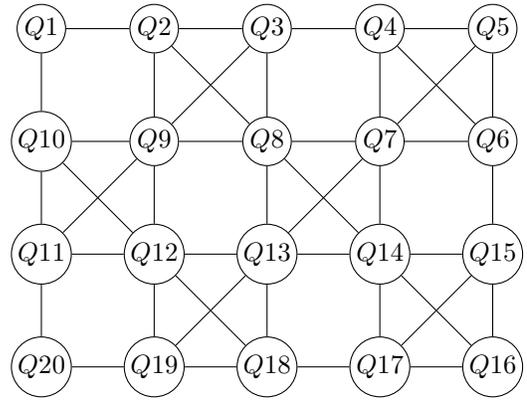

		\ctikzfig{ibm_q20_tokyo}
		\caption{ibm-q20-tokyo} \label{ibm_q20_tokyo}
	\end{figure}
	
$$M_0=\begin{cases}
w_1 \mapsto Q1, & w_2 \mapsto Q2 \\
w_3 \mapsto Q3, & w_4 \mapsto Q4 \\
w_5 \mapsto Q5 , & w_6 \mapsto Q6 \\
w_7 \mapsto Q7 , & w_8 \mapsto Q8 \\
w_9 \mapsto Q9 , & w_{10} \mapsto Q10 \\
w_{11} \mapsto Q11 , & w_{12} \mapsto Q12 \\
w_{13} \mapsto Q13 , & w_{14} \mapsto Q14 \\
w_{15} \mapsto Q15 , & w_{16} \mapsto Q16 \\
w_{17} \mapsto Q17 , & w_{18} \mapsto Q18 \\
w_{19} \mapsto Q19 , & w_{20} \mapsto Q20 \\
\end{cases}$$

\item ibm-qx5:\\
	See Figure \ref{fig: ibmqx5}.
 	\begin{figure*}[t]
		\ctikzfig{ibmqx5}
		\caption{ibm-qx5} \label{fig: ibmqx5}
	\end{figure*}
	
$$M_0=\begin{cases}
w_1 \mapsto Q1, & w_2 \mapsto Q2 \\
w_3 \mapsto Q3, & w_4 \mapsto Q4 \\
w_5 \mapsto Q5 , & w_6 \mapsto Q6 \\
w_7 \mapsto Q7 , & w_8 \mapsto Q8 \\
w_9 \mapsto Q9 , & w_{10} \mapsto Q10 \\
w_{11} \mapsto Q11 , & w_{12} \mapsto Q12 \\
w_{13} \mapsto Q13 , & w_{14} \mapsto Q14 \\
w_{15} \mapsto Q15 , & w_{16} \mapsto Q16 \\
\end{cases}$$

\end{itemize}

\end{document}

%% file: sample.tikzstyles
\tikzstyle{output}=[fill={rgb,255: red,19; green,212; blue,255}, draw=black, shape=circle, tikzit fill={rgb,255: red,19; green,212; blue,255}, tikzit draw={rgb,255: red,191; green,191; blue,191}, inner sep=0pt, minimum size=10pt]
\tikzstyle{input}=[fill={rgb,255: red,199; green,255; blue,254}, draw=black, shape=circle, tikzit fill={rgb,255: red,199; green,255; blue,254}, tikzit draw=black, inner sep=0pt, minimum size=10pt]
\tikzstyle{state-plain}=[fill=white, draw=black, tikzit shape=rectangle, draw, shape border uses incircle, isosceles triangle, isosceles triangle apex angle=60, shape border rotate=-90, inner sep=0pt, minimum height=1.5em]
\tikzstyle{effect-plain}=[fill=white, draw=black, tikzit shape=rectangle, draw, shape border uses incircle, isosceles triangle, isosceles triangle apex angle=60, shape border rotate=90, inner sep=0pt, minimum height=1.5em]
\tikzstyle{z-single}=[fill=green, draw=black, shape=circle, tikzit fill=green, tikzit draw={rgb,255: red,191; green,191; blue,191}, inner sep=0pt, minimum size=10pt]
\tikzstyle{x-single}=[fill=red, draw=black, shape=circle, tikzit fill=red, tikzit draw={rgb,255: red,191; green,191; blue,191}, inner sep=0pt, minimum size=10pt]
\tikzstyle{z-double}=[fill=green, ultra thick, draw=black, shape=circle]
\tikzstyle{x-double}=[fill=red, ultra thick, draw=black, shape=circle]
\tikzstyle{box}=[fill=white, draw=black, shape=rectangle]
\tikzstyle{large-box}=[fill=white, draw=black, shape=rectangle, minimum width=1.5cm]
\tikzstyle{medium-box}=[fill=white, draw=black, shape=rectangle, minimum width=1.0cm]
\tikzstyle{trap-double}=[fill=white, draw=black, trapezium, trapezium left angle=60, trapezium right angle=0, ultra thick, minimum width=1cm, minimum height=1cm, inner ysep=0pt, inner xsep=3pt, tikzit shape=rectangle]
\tikzstyle{trap*-double}=[fill=white, draw=black, trapezium, trapezium left angle=-60, trapezium right angle=0, ultra thick, minimum width=1cm, minimum height=1cm, inner ysep=0pt, tikzit shape=rectangle, inner xsep=3pt]
\tikzstyle{trapR-double}=[fill=white, draw=black, trapezium, trapezium left angle=60, trapezium right angle=0, ultra thick, minimum width=1cm, minimum height=1cm, inner ysep=0pt, inner xsep=3pt, tikzit shape=rectangle]
\tikzstyle{trapR*-double}=[fill=white, draw=black, trapezium, trapezium left angle=0, trapezium right angle=60, ultra thick, minimum width=1cm, minimum height=1cm, inner ysep=0pt, tikzit shape=rectangle, inner xsep=3pt]
\tikzstyle{state-z}=[fill=green, draw=black, tikzit shape=rectangle, draw, shape border uses incircle, isosceles triangle, isosceles triangle apex angle=60, shape border rotate=-90, inner sep=0pt, minimum height=2em]
\tikzstyle{state-x}=[fill=red, draw=black, tikzit shape=rectangle, draw, shape border uses incircle, isosceles triangle, isosceles triangle apex angle=60, shape border rotate=-90, inner sep=0pt, minimum height=2em]
\tikzstyle{x-box}=[fill=red, draw=black, shape=rectangle]
\tikzstyle{effect-z}=[fill=green, draw=black, tikzit shape=rectangle, draw, shape border uses incircle, isosceles triangle, isosceles triangle apex angle=60, shape border rotate=90, inner sep=0pt, minimum height=2em]
\tikzstyle{effect-x}=[fill=red, draw=black, tikzit shape=rectangle, draw, shape border uses incircle, isosceles triangle, isosceles triangle apex angle=60, shape border rotate=90, inner sep=0pt, minimum height=2em]
\tikzstyle{H}=[fill=yellow, draw=black, shape=rectangle]
\tikzstyle{dot}=[fill=black, draw=black, shape=circle, inner sep=1pt]
\tikzstyle{white-dot}=[fill=white, draw=black, shape=circle, inner sep=1pt]
\tikzstyle{pg}=[fill={rgb,255: red,255; green,128; blue,0}, draw=yellow, shape=rectangle, inner sep=2pt]
\tikzstyle{ntpg}=[fill=cyan, draw={rgb,255: red,0; green,128; blue,128}, shape=rectangle, inner sep=2pt]
\tikzstyle{z-ht}=[fill=green, draw=magenta, shape=circle, inner sep=0pt, minimum size=10pt]
\tikzstyle{z-ht2}=[fill={rgb,255: red,191; green,255; blue,0}, draw=magenta, shape=circle, inner sep=0pt, minimum size=10pt]
\tikzstyle{z-ht3}=[fill={rgb,255: red,191; green,255; blue,0}, draw=black, shape=circle, inner sep=0pt, minimum size=10pt]
\tikzstyle{pi}=[fill={rgb,255: red,0; green,128; blue,128}, draw=black, shape=rectangle, inner sep=2pt]
\tikzstyle{halfpi}=[fill={rgb,255: red,255; green,191; blue,191}, draw=black, shape=rectangle, inner sep=2pt]
\tikzstyle{qpi}=[fill=blue, draw=black, shape=rectangle, inner sep=2pt]
\tikzstyle{cross}=[path picture={\draw[thick,black](path picture bounding box.north) -- (path picture bounding box.south) (path picture bounding box.west) -- (path picture bounding box.east);
}]
\tikzstyle{swap}=[path picture={\draw[thick,black,inner sep=0pt](path picture bounding box.south east) -- (path picture bounding box.north west) (path picture bounding box.south west) -- (path picture bounding box.north east);
}]
\tikzstyle{control}=[draw, fill, shape=circle, minimum size=5pt, inner sep=0pt]
\tikzstyle{target}=[draw=black, shape=circle, cross, minimum width=0.3 cm, tikzit draw=black, tikzit fill={rgb,255: red,191; green,191; blue,191}]

\tikzstyle{single}=[-, tikzit draw={rgb,255: red,128; green,128; blue,0}, draw=black]
\tikzstyle{double}=[-, draw=black, ultra thick, tikzit draw=black]
\tikzstyle{Hline}=[-, draw={rgb,255: red,47; green,141; blue,255}, dashed]
\tikzstyle{pgl}=[-, draw=magenta]
\tikzstyle{arrowup}=[->]
\tikzstyle{arrowdown}=[<-]
\tikzstyle{dotted}=[-, draw={rgb,255: red,128; green,128; blue,128}, dashed]
\tikzstyle{connection}=[-, dashed, draw=green]
\tikzstyle{red}=[-, draw=red]
\tikzstyle{double blue}=[-, ultra thick, draw=blue]

%% file: figures/H.tikz
\begin{tikzpicture}
	\begin{pgfonlayer}{nodelayer}
		\node [style=none] (0) at (-2, 0) {};
		\node [style=none] (1) at (2, 0) {};
		\node [style=box] (2) at (0, 0) {\textbf{H}};
	\end{pgfonlayer}
	\begin{pgfonlayer}{edgelayer}
		\draw [style=single] (0.center) to (1.center);
	\end{pgfonlayer}
\end{tikzpicture}

%% file: figures/cnot.tikz
\begin{tikzpicture}
	\begin{pgfonlayer}{nodelayer}
		\node [style=none] (0) at (-2, 1) {};
		\node [style=none] (1) at (2, 1) {};
		\node [style=none] (2) at (-2, -1) {};
		\node [style=none] (3) at (2, -1) {};
		\node [style=control] (4) at (0, 1) {};
		\node [style=target] (5) at (0, -1) {};
	\end{pgfonlayer}
	\begin{pgfonlayer}{edgelayer}
		\draw [style=single] (0.center) to (1.center);
		\draw [style=single] (2.center) to (3.center);
		\draw [style=single] (4) to (5);
	\end{pgfonlayer}
\end{tikzpicture}

%% file: figures/cnot_reverse.tikz
\begin{tikzpicture}
	\begin{pgfonlayer}{nodelayer}
		\node [style=none] (0) at (-3.25, 1) {};
		\node [style=none] (1) at (3.25, 1) {};
		\node [style=none] (2) at (-3.25, -1) {};
		\node [style=none] (3) at (3.25, -1) {};
		\node [style=control] (4) at (0, -1) {};
		\node [style=target] (5) at (0, 1) {};
		\node [style=box] (6) at (-2, 1) {\textbf{H}};
		\node [style=box] (7) at (2, -1) {\textbf{H}};
		\node [style=box] (8) at (2, 1) {\textbf{H}};
		\node [style=box] (9) at (-2, -1) {\textbf{H}};
	\end{pgfonlayer}
	\begin{pgfonlayer}{edgelayer}
		\draw [style=single] (0.center) to (1.center);
		\draw [style=single] (2.center) to (3.center);
		\draw [style=single] (4) to (5);
	\end{pgfonlayer}
\end{tikzpicture}

%% file: figures/ex_routing_1_c.tikz
\begin{tikzpicture}
	\begin{pgfonlayer}{nodelayer}
		\node [style=none] (0) at (-5, 3) {};
		\node [style=none] (1) at (3.75, 3) {};
		\node [style=none] (2) at (-5, 1) {};
		\node [style=none] (3) at (3.75, 1) {};
		\node [style=control] (4) at (-1, 1) {};
		\node [style=target] (5) at (-1, 3) {};
		\node [style=box] (6) at (1.75, 3) {\textbf{X}};
		\node [style=none] (10) at (-5, -1) {};
		\node [style=none] (11) at (3.75, -1) {};
		\node [style=none] (12) at (-5, -3) {};
		\node [style=none] (13) at (3.75, -3) {};
		\node [style=control] (14) at (0.75, 1) {};
		\node [style=target] (15) at (0.75, -3) {};
		\node [style=box] (16) at (-2, -1) {\textbf{H}};
		\node [style=none] (17) at (-7, 3) {$w_1$};
		\node [style=none] (18) at (-7, 1) {$w_2$};
		\node [style=none] (19) at (-7, -1) {$w_3$};
		\node [style=none] (20) at (-7, -3) {$w_4$};
	\end{pgfonlayer}
	\begin{pgfonlayer}{edgelayer}
		\draw [style=single] (0.center) to (1.center);
		\draw [style=single] (2.center) to (3.center);
		\draw [style=single] (4) to (5);
		\draw [style=single] (10.center) to (11.center);
		\draw [style=single] (12.center) to (13.center);
		\draw [style=single] (14) to (15);
	\end{pgfonlayer}
\end{tikzpicture}

%% file: figures/ex_routing_1_g.tikz
\begin{tikzpicture}
	\begin{pgfonlayer}{nodelayer}
		\node [style=white-dot] (4) at (-2, 2) {$Q1$};
		\node [style=white-dot] (5) at (2, 2) {$Q2$};
		\node [style=white-dot] (6) at (-2, -2) {$Q3$};
		\node [style=white-dot] (7) at (2, -2) {$Q4$};
	\end{pgfonlayer}
	\begin{pgfonlayer}{edgelayer}
		\draw [style=single] (4) to (5);
		\draw [style=single] (5) to (7);
		\draw [style=single] (7) to (6);
		\draw [style=single] (6) to (4);
	\end{pgfonlayer}
\end{tikzpicture}

%% file: figures/X.tikz
\begin{tikzpicture}
	\begin{pgfonlayer}{nodelayer}
		\node [style=none] (0) at (-2, 0) {};
		\node [style=none] (1) at (2, 0) {};
		\node [style=box] (2) at (0, 0) {\textbf{X}};
	\end{pgfonlayer}
	\begin{pgfonlayer}{edgelayer}
		\draw [style=single] (0.center) to (1.center);
	\end{pgfonlayer}
\end{tikzpicture}

%% file: figures/SWAP.tikz
\begin{tikzpicture}
	\begin{pgfonlayer}{nodelayer}
		\node [style=none] (0) at (-2, 1) {};
		\node [style=none] (1) at (2, 1) {};
		\node [style=none] (2) at (-2, -1) {};
		\node [style=none] (3) at (2, -1) {};
		\node [style=swap] (4) at (0, 1) {};
		\node [style=swap] (5) at (0, -1) {};
	\end{pgfonlayer}
	\begin{pgfonlayer}{edgelayer}
		\draw [style=single] (0.center) to (1.center);
		\draw [style=single] (2.center) to (3.center);
		\draw [style=single] (4) to (5);
	\end{pgfonlayer}
\end{tikzpicture}

%% file: figures/CZ.tikz
\begin{tikzpicture}
	\begin{pgfonlayer}{nodelayer}
		\node [style=none] (0) at (-2, 1) {};
		\node [style=none] (1) at (2, 1) {};
		\node [style=none] (2) at (-2, -1) {};
		\node [style=none] (3) at (2, -1) {};
		\node [style=control] (4) at (0, 1) {};
		\node [style=control] (5) at (0, -1) {};
	\end{pgfonlayer}
	\begin{pgfonlayer}{edgelayer}
		\draw [style=single] (0.center) to (1.center);
		\draw [style=single] (2.center) to (3.center);
		\draw [style=single] (4) to (5);
	\end{pgfonlayer}
\end{tikzpicture}

%% file: figures/CZ_3.tikz
\begin{tikzpicture}
	\begin{pgfonlayer}{nodelayer}
		\node [style=none] (0) at (-2.5, 1) {};
		\node [style=none] (1) at (-2.5, -1) {};
		\node [style=none] (2) at (2.5, 1) {};
		\node [style=none] (3) at (2.5, -1) {};
		\node [style=target] (4) at (0, 1) {};
		\node [style=control] (5) at (0, -1) {};
		\node [style=box] (6) at (1.5, 1) {$H$};
		\node [style=box] (7) at (-1.5, 1) {$H$};
	\end{pgfonlayer}
	\begin{pgfonlayer}{edgelayer}
		\draw [style=single] (0.center) to (2.center);
		\draw [style=single] (1.center) to (3.center);
		\draw [style=single] (4) to (5);
	\end{pgfonlayer}
\end{tikzpicture}

%% file: figures/SWAP_CNOT.tikz
\begin{tikzpicture}
	\begin{pgfonlayer}{nodelayer}
		\node [style=none] (0) at (-3, 1) {};
		\node [style=none] (1) at (-3, -1) {};
		\node [style=none] (2) at (3, 1) {};
		\node [style=none] (3) at (3, -1) {};
		\node [style=control] (4) at (-2, -1) {};
		\node [style=target] (5) at (-2, 1) {};
		\node [style=target] (6) at (0, -1) {};
		\node [style=control] (7) at (0, 1) {};
		\node [style=control] (8) at (2, -1) {};
		\node [style=target] (9) at (2, 1) {};
	\end{pgfonlayer}
	\begin{pgfonlayer}{edgelayer}
		\draw [style=single] (0.center) to (2.center);
		\draw [style=single] (1.center) to (3.center);
		\draw [style=single] (9) to (8);
		\draw [style=single] (7) to (6);
		\draw [style=single] (5) to (4);
	\end{pgfonlayer}
\end{tikzpicture}

%% file: figures/SWAP_CNOT2.tikz
\begin{tikzpicture}
	\begin{pgfonlayer}{nodelayer}
		\node [style=none] (0) at (-3, 1) {};
		\node [style=none] (1) at (-3, -1) {};
		\node [style=none] (2) at (3, 1) {};
		\node [style=none] (3) at (3, -1) {};
		\node [style=target] (6) at (-2, -1) {};
		\node [style=control] (7) at (-2, 1) {};
		\node [style=control] (8) at (0, -1) {};
		\node [style=target] (9) at (0, 1) {};
		\node [style=target] (10) at (2, -1) {};
		\node [style=control] (11) at (2, 1) {};
	\end{pgfonlayer}
	\begin{pgfonlayer}{edgelayer}
		\draw [style=single] (0.center) to (2.center);
		\draw [style=single] (1.center) to (3.center);
		\draw [style=single] (9) to (8);
		\draw [style=single] (7) to (6);
		\draw [style=single] (11) to (10);
	\end{pgfonlayer}
\end{tikzpicture}

%% file: figures/CNOT_2.tikz
\begin{tikzpicture}
	\begin{pgfonlayer}{nodelayer}
		\node [style=none] (0) at (-3, 2) {};
		\node [style=none] (1) at (-3, 0) {};
		\node [style=none] (2) at (-3, -2) {};
		\node [style=none] (3) at (3, 2) {};
		\node [style=none] (4) at (3, 0) {};
		\node [style=none] (5) at (3, -2) {};
		\node [style=target] (6) at (0, -2) {};
		\node [style=control] (7) at (0, 2) {};
	\end{pgfonlayer}
	\begin{pgfonlayer}{edgelayer}
		\draw [style=single] (0.center) to (3.center);
		\draw [style=single] (1.center) to (4.center);
		\draw [style=single] (2.center) to (5.center);
		\draw [style=single] (7) to (6);
	\end{pgfonlayer}
\end{tikzpicture}

%% file: figures/double_cnot.tikz
\begin{tikzpicture}
	\begin{pgfonlayer}{nodelayer}
		\node [style=none] (0) at (-3, 1) {};
		\node [style=none] (1) at (3, 1) {};
		\node [style=none] (2) at (-3, -1) {};
		\node [style=none] (3) at (3, -1) {};
		\node [style=target] (4) at (-1, -1) {};
		\node [style=control] (5) at (-1, 1) {};
		\node [style=control] (6) at (1, 1) {};
		\node [style=target] (7) at (1, -1) {};
	\end{pgfonlayer}
	\begin{pgfonlayer}{edgelayer}
		\draw [style=single] (0.center) to (1.center);
		\draw [style=single] (2.center) to (3.center);
		\draw [style=single] (6) to (7);
		\draw [style=single] (5) to (4);
	\end{pgfonlayer}
\end{tikzpicture}

%% file: figures/Identity.tikz
\begin{tikzpicture}
	\begin{pgfonlayer}{nodelayer}
		\node [style=none] (0) at (-3, 1) {};
		\node [style=none] (1) at (3, 1) {};
		\node [style=none] (2) at (-3, -1) {};
		\node [style=none] (3) at (3, -1) {};
	\end{pgfonlayer}
	\begin{pgfonlayer}{edgelayer}
		\draw [style=single] (0.center) to (1.center);
		\draw [style=single] (2.center) to (3.center);
	\end{pgfonlayer}
\end{tikzpicture}

%% file: figures/synthesis_permutation_c.tikz
\begin{tikzpicture}
	\begin{pgfonlayer}{nodelayer}
		\node [style=none] (0) at (-4, 3) {};
		\node [style=none] (1) at (4, 3) {};
		\node [style=none] (2) at (-4, 1.5) {};
		\node [style=none] (3) at (4, 1.5) {};
		\node [style=none] (4) at (-4, 0) {};
		\node [style=none] (5) at (4, 0) {};
		\node [style=none] (6) at (-4, -1.5) {};
		\node [style=none] (7) at (4, -1.5) {};
		\node [style=control] (8) at (-2, 3) {};
		\node [style=target] (9) at (-2, 0) {};
		\node [style=control] (10) at (0, 0) {};
		\node [style=target] (11) at (0, -1.5) {};
		\node [style=none] (12) at (-5.25, 3) {$w_0$};
		\node [style=none] (13) at (-5.25, 1.5) {$w_1$};
		\node [style=none] (14) at (-5.25, 0) {$w_2$};
		\node [style=none] (15) at (-5.25, -1.5) {$w_3$};
	\end{pgfonlayer}
	\begin{pgfonlayer}{edgelayer}
		\draw [style=single] (0.center) to (1.center);
		\draw [style=single] (2.center) to (3.center);
		\draw [style=single] (4.center) to (5.center);
		\draw [style=single] (6.center) to (7.center);
		\draw [style=single] (8) to (9);
		\draw [style=single] (10) to (11);
	\end{pgfonlayer}
\end{tikzpicture}

%% file: figures/4_grid.tikz
\begin{tikzpicture}
	\begin{pgfonlayer}{nodelayer}
		\node [style=white-dot] (0) at (-2, 2) {$A$};
		\node [style=white-dot] (1) at (2, 2) {$B$};
		\node [style=white-dot] (2) at (2, -2) {$C$};
		\node [style=white-dot] (3) at (-2, -2) {$D$};
	\end{pgfonlayer}
	\begin{pgfonlayer}{edgelayer}
		\draw [style=single] (0) to (1);
		\draw [style=single] (1) to (2);
		\draw [style=single] (2) to (3);
		\draw [style=single] (0) to (3);
	\end{pgfonlayer}
\end{tikzpicture}

%% file: figures/matrix_graph.tikz
\begin{tikzpicture}
	\begin{pgfonlayer}{nodelayer}
		\node [style=white-dot] (0) at (2, 2) {$A$};
		\node [style=white-dot] (1) at (6, 2) {$B$};
		\node [style=white-dot] (2) at (6, -2) {$C$};
		\node [style=white-dot] (3) at (2, -2) {$D$};
		\node [style=none] (4) at (0.5, 3.5) {\small (1 0 1 1)};
		\node [style=none] (5) at (7.5, 3.5) {\small (0 1 0 0)};
		\node [style=none] (6) at (7.5, -3.5) {\small (0 0 1 1)};
		\node [style=none] (7) at (0.5, -3.5) {\small (0 0 0 1)};
	\end{pgfonlayer}
	\begin{pgfonlayer}{edgelayer}
		\draw [style=single] (0) to (1);
		\draw [style=single] (1) to (2);
		\draw [style=single] (2) to (3);
		\draw [style=single] (0) to (3);
	\end{pgfonlayer}
\end{tikzpicture}

%% file: figures/matrix_graph_swap.tikz
\begin{tikzpicture}
	\begin{pgfonlayer}{nodelayer}
		\node [style=white-dot] (0) at (2, 2) {$A$};
		\node [style=white-dot] (1) at (6, 2) {$B$};
		\node [style=white-dot] (2) at (6, -2) {$C$};
		\node [style=white-dot] (3) at (2, -2) {$D$};
		\node [style=none] (4) at (7.5, 3.5) {\color{red} \small (1 0 1 1)};
		\node [style=none] (5) at (0.5, 3.5) {\color{red} \small (0 1 0 0)};
		\node [style=none] (6) at (7.5, -3.5) {\small (0 0 1 1)};
		\node [style=none] (7) at (0.5, -3.5) {\small (0 0 0 1)};
	\end{pgfonlayer}
	\begin{pgfonlayer}{edgelayer}
		\draw [style=single] (0) to (1);
		\draw [style=single] (1) to (2);
		\draw [style=single] (2) to (3);
		\draw [style=single] (0) to (3);
	\end{pgfonlayer}
\end{tikzpicture}

%% file: figures/matrix_graph_B_C.tikz
\begin{tikzpicture}
	\begin{pgfonlayer}{nodelayer}
		\node [style=white-dot] (0) at (-2, 2) {$A$};
		\node [style=white-dot] (1) at (2, 2) {$B$};
		\node [style=white-dot] (2) at (2, -2) {$C$};
		\node [style=white-dot] (3) at (-2, -2) {$D$};
		\node [style=none] (4) at (3.5, 3.5) {\color{red} \small (1 0 0 0)};
		\node [style=none] (5) at (-3.5, 3.5) {\small (0 1 0 0)};
		\node [style=none] (6) at (3.5, -3.5) {\small (0 0 1 1)};
		\node [style=none] (7) at (-3.5, -3.5) {\small (0 0 0 1)};
	\end{pgfonlayer}
	\begin{pgfonlayer}{edgelayer}
		\draw [style=single] (0) to (1);
		\draw [style=single] (1) to (2);
		\draw [style=single] (2) to (3);
		\draw [style=single] (0) to (3);
	\end{pgfonlayer}
\end{tikzpicture}

%% file: figures/matrix_graph_C_D.tikz
\begin{tikzpicture}
	\begin{pgfonlayer}{nodelayer}
		\node [style=white-dot] (0) at (-2, 2) {$A$};
		\node [style=white-dot] (1) at (2, 2) {$B$};
		\node [style=white-dot] (2) at (2, -2) {$C$};
		\node [style=white-dot] (3) at (-2, -2) {$D$};
		\node [style=none] (4) at (3.5, 3.5) {\small (1 0 0 0)};
		\node [style=none] (5) at (-3.5, 3.5) {\small (0 1 0 0)};
		\node [style=none] (6) at (3.5, -3.5) {\color{red} \small (0 0 1 0)};
		\node [style=none] (7) at (-3.5, -3.5) {\small (0 0 0 1)};
	\end{pgfonlayer}
	\begin{pgfonlayer}{edgelayer}
		\draw [style=single] (0) to (1);
		\draw [style=single] (1) to (2);
		\draw [style=single] (2) to (3);
		\draw [style=single] (0) to (3);
	\end{pgfonlayer}
\end{tikzpicture}

%% file: figures/synthesis_permutation_c_p.tikz
\begin{tikzpicture}
	\begin{pgfonlayer}{nodelayer}
		\node [style=none] (0) at (-3.25, 2.5) {};
		\node [style=none] (1) at (4.75, 2.5) {};
		\node [style=none] (2) at (-3.25, 1) {};
		\node [style=none] (3) at (4.75, 1) {};
		\node [style=none] (4) at (-3.25, -0.5) {};
		\node [style=none] (5) at (4.75, -0.5) {};
		\node [style=none] (6) at (-3.25, -2) {};
		\node [style=none] (7) at (4.75, -2) {};
		\node [style=control] (8) at (-2, 2.5) {};
		\node [style=target] (9) at (-2, 1) {};
		\node [style=control] (10) at (2.5, 1) {};
		\node [style=target] (11) at (2.5, -0.5) {};
		\node [style=none] (12) at (-4.5, 2.5) {$w_0$};
		\node [style=none] (13) at (-4.5, 1) {$w_1$};
		\node [style=none] (14) at (-4.5, -0.5) {$w_2$};
		\node [style=none] (15) at (-4.5, -2) {$w_3$};
		\node [style=control] (16) at (-0.5, 1) {};
		\node [style=target] (17) at (-0.5, 2.5) {};
		\node [style=control] (18) at (1, 2.5) {};
		\node [style=target] (19) at (1, 1) {};
		\node [style=target] (20) at (4, -2) {};
		\node [style=control] (21) at (4, -0.5) {};
	\end{pgfonlayer}
	\begin{pgfonlayer}{edgelayer}
		\draw [style=single] (0.center) to (1.center);
		\draw [style=single] (2.center) to (3.center);
		\draw [style=single] (4.center) to (5.center);
		\draw [style=single] (6.center) to (7.center);
		\draw [style=single] (8) to (9);
		\draw [style=single] (10) to (11);
		\draw [style=single] (17) to (16);
		\draw [style=single] (18) to (19);
		\draw [style=single] (21) to (20);
	\end{pgfonlayer}
\end{tikzpicture}

%% file: figures/tree_reduction.tikz
\begin{tikzpicture}
	\begin{pgfonlayer}{nodelayer}
		\node [style=white-dot] (0) at (0, 4) {$R$};
		\node [style=input] (1) at (-2, -0.5) {$C$};
		\node [style=white-dot] (2) at (2, 2) {$B$};
		\node [style=white-dot] (3) at (-3.5, -2.5) {$D$};
		\node [style=white-dot] (4) at (-0.5, -2.5) {$E$};
		\node [style=none] (7) at (0, 5.5) {\small $e_1+e_2+e_3$};
		\node [style=none] (8) at (3.5, 2) {\small $e_1$};
		\node [style=none] (9) at (-4.25, -4) {\small $e_2$};
		\node [style=none] (10) at (0.25, -4) {\small $e_3+e_4$};
		\node [style=none] (13) at (-4, -0.5) {\small $e_7+e_5$};
		\node [style=input] (14) at (-2, 2) {$A$};
		\node [style=none] (15) at (-4, 2) {\small $e_6$};
		\node [style=none] (16) at (-5.5, -2.5) {\color{orange} $\to$};
	\end{pgfonlayer}
	\begin{pgfonlayer}{edgelayer}
		\draw [style=single] (1) to (3);
		\draw [style=single] (1) to (4);
		\draw [style=single] (0) to (2);
		\draw [style=single] (0) to (14);
		\draw [style=single] (14) to (1);
	\end{pgfonlayer}
\end{tikzpicture}

%% file: figures/tree_reduction_2.tikz
\begin{tikzpicture}
	\begin{pgfonlayer}{nodelayer}
		\node [style=white-dot] (0) at (0, 4) {$R$};
		\node [style=white-dot] (1) at (-2, -0.5) {$C$};
		\node [style=white-dot] (2) at (2, 2) {$B$};
		\node [style=input] (3) at (-3.5, -2.5) {$D$};
		\node [style=white-dot] (4) at (-0.5, -2.5) {$E$};
		\node [style=none] (5) at (0, 5.5) {\small $e_1+e_2+e_3$};
		\node [style=none] (6) at (3.5, 2) {\small $e_1$};
		\node [style=none] (7) at (-4, -0.5) {\color{red} \small $e_2$};
		\node [style=none] (8) at (0.25, -4) {\small $e_3+e_4$};
		\node [style=none] (9) at (-4.5, -4) {\color{red} \small $e_7+e_5$};
		\node [style=input] (10) at (-2, 2) {$A$};
		\node [style=none] (11) at (-4, 2) {\small $e_6$};
		\node [style=none] (13) at (2, -2.5) {\color{orange} $\gets$};
	\end{pgfonlayer}
	\begin{pgfonlayer}{edgelayer}
		\draw [style=single] (1) to (3);
		\draw [style=single] (1) to (4);
		\draw [style=single] (0) to (2);
		\draw [style=single] (0) to (10);
		\draw [style=single] (10) to (1);
	\end{pgfonlayer}
\end{tikzpicture}

%% file: figures/tree_reduction_3.tikz
\begin{tikzpicture}
	\begin{pgfonlayer}{nodelayer}
		\node [style=white-dot] (0) at (0, 4) {$R$};
		\node [style=white-dot] (1) at (-2, -0.5) {$C$};
		\node [style=white-dot] (2) at (2, 2) {$B$};
		\node [style=input] (3) at (-3.5, -2.5) {$D$};
		\node [style=white-dot] (4) at (-0.5, -2.5) {$E$};
		\node [style=none] (5) at (0, 5.5) {\small $e_1+e_2+e_3$};
		\node [style=none] (6) at (3.5, 2) {\small $e_1$};
		\node [style=none] (7) at (-5, -0.5) {\color{red} \small $e_2+e_3+e_4$};
		\node [style=none] (8) at (0.5, -4) {\small $e_3+e_4$};
		\node [style=none] (9) at (-4.5, -4) {\small $e_7+e_5$};
		\node [style=input] (10) at (-2, 2) {$A$};
		\node [style=none] (11) at (-4, 2) {\small $e_6$};
		\node [style=none] (13) at (0.5, -0.5) {\color{orange} $\gets$};
	\end{pgfonlayer}
	\begin{pgfonlayer}{edgelayer}
		\draw [style=single] (1) to (3);
		\draw [style=single] (1) to (4);
		\draw [style=single] (0) to (2);
		\draw [style=single] (0) to (10);
		\draw [style=single] (10) to (1);
	\end{pgfonlayer}
\end{tikzpicture}

%% file: figures/tree_reduction_6.tikz
\begin{tikzpicture}
	\begin{pgfonlayer}{nodelayer}
		\node [style=white-dot] (0) at (0, 4) {$R$};
		\node [style=input] (1) at (-2, -0.5) {$C$};
		\node [style=white-dot] (2) at (2, 2) {$B$};
		\node [style=input] (3) at (-3.5, -2.5) {$D$};
		\node [style=white-dot] (4) at (-0.5, -2.5) {$E$};
		\node [style=none] (5) at (0, 5.5) {\color{red} \small $e_4$};
		\node [style=none] (6) at (3.5, 2) {\small $e_1$};
		\node [style=none] (7) at (-5.25, 2) {\small $e_2+e_3+e_4$};
		\node [style=none] (8) at (0.5, -4) {\small $e_3+e_4$};
		\node [style=none] (9) at (-4.5, -4) {\small $e_7+e_5$};
		\node [style=white-dot] (10) at (-2, 2) {$A$};
		\node [style=none] (11) at (-4, -0.5) {\small $e_6$};
		\node [style=none] (12) at (-1.5, 4) {\color{orange} $\to$};
	\end{pgfonlayer}
	\begin{pgfonlayer}{edgelayer}
		\draw [style=single] (1) to (3);
		\draw [style=single] (1) to (4);
		\draw [style=single] (0) to (2);
		\draw [style=single] (0) to (10);
		\draw [style=single] (10) to (1);
	\end{pgfonlayer}
\end{tikzpicture}

%% file: figures/tree_reduction_r_0.tikz
\begin{tikzpicture}
	\begin{pgfonlayer}{nodelayer}
		\node [style=white-dot] (0) at (0, 4) {$R$};
		\node [style=input] (1) at (-2, -0.5) {$C$};
		\node [style=white-dot] (2) at (2, 2) {$B$};
		\node [style=input] (3) at (-3.5, -2.5) {$D$};
		\node [style=white-dot] (4) at (-0.5, -2.5) {$E$};
		\node [style=none] (5) at (0, 5.5) {\color{red} \small $e_4$};
		\node [style=none] (6) at (3.5, 2) {\small $e_1$};
		\node [style=none] (7) at (-5.25, 2) {\small $e_2+e_3+e_4$};
		\node [style=none] (8) at (0.5, -4) {\small $e_3+e_4$};
		\node [style=none] (9) at (-4.5, -4) {\small $e_7+e_5$};
		\node [style=white-dot] (10) at (-2, 2) {$A$};
		\node [style=none] (11) at (-4, -0.5) {\small $e_6$};
	\end{pgfonlayer}
	\begin{pgfonlayer}{edgelayer}
		\draw [style=single] (1) to (3);
		\draw [style=single] (1) to (4);
		\draw [style=single] (0) to (2);
		\draw [style=single] (0) to (10);
		\draw [style=single] (10) to (1);
	\end{pgfonlayer}
\end{tikzpicture}

%% file: figures/tree_reduction_r_1.tikz
\begin{tikzpicture}
	\begin{pgfonlayer}{nodelayer}
		\node [style=white-dot] (0) at (0, 4) {$R$};
		\node [style=white-dot] (1) at (-2, -0.5) {$C$};
		\node [style=white-dot] (2) at (2, 2) {$B$};
		\node [style=input] (3) at (-3.5, -2.5) {$D$};
		\node [style=white-dot] (4) at (-0.5, -2.5) {$E$};
		\node [style=none] (5) at (0, 5.5) {\small $e_4$};
		\node [style=none] (6) at (3.5, 2) {\small $e_1$};
		\node [style=none] (7) at (-4.75, -0.5) {\color{red} \small $e_2+e_3+e_4$};
		\node [style=none] (8) at (0.5, -4) {\small $e_3+e_4$};
		\node [style=none] (9) at (-4.5, -4) {\small $e_7+e_5$};
		\node [style=input] (10) at (-2, 2) {$A$};
		\node [style=none] (11) at (-5.25, 2) {\color{red} \small $e_6$};
	\end{pgfonlayer}
	\begin{pgfonlayer}{edgelayer}
		\draw [style=single] (1) to (3);
		\draw [style=single] (1) to (4);
		\draw [style=single] (0) to (2);
		\draw [style=single] (0) to (10);
		\draw [style=single] (10) to (1);
	\end{pgfonlayer}
\end{tikzpicture}

%% file: figures/tree_reduction_r_2.tikz
\begin{tikzpicture}
	\begin{pgfonlayer}{nodelayer}
		\node [style=white-dot] (0) at (0, 4) {$R$};
		\node [style=white-dot] (1) at (-2, -0.5) {$C$};
		\node [style=white-dot] (2) at (2, 2) {$B$};
		\node [style=input] (3) at (-3.5, -2.5) {$D$};
		\node [style=white-dot] (4) at (-0.5, -2.5) {$E$};
		\node [style=none] (5) at (0, 5.5) {\small $e_4$};
		\node [style=none] (6) at (3.5, 2) {\small $e_1$};
		\node [style=none] (7) at (-4.75, -0.5) {\color{red} \small $e_2$};
		\node [style=none] (8) at (0.5, -4) {\small $e_3+e_4$};
		\node [style=none] (9) at (-4.5, -4) {\small $e_7+e_5$};
		\node [style=input] (10) at (-2, 2) {$A$};
		\node [style=none] (11) at (-5.25, 2) {\small $e_6$};
	\end{pgfonlayer}
	\begin{pgfonlayer}{edgelayer}
		\draw [style=single] (1) to (3);
		\draw [style=single] (1) to (4);
		\draw [style=single] (0) to (2);
		\draw [style=single] (0) to (10);
		\draw [style=single] (10) to (1);
	\end{pgfonlayer}
\end{tikzpicture}

%% file: figures/tree_reduction_r_3.tikz
\begin{tikzpicture}
	\begin{pgfonlayer}{nodelayer}
		\node [style=white-dot] (0) at (0, 4) {$R$};
		\node [style=input] (1) at (-2, -0.5) {$C$};
		\node [style=white-dot] (2) at (2, 2) {$B$};
		\node [style=white-dot] (3) at (-3.5, -2.5) {$D$};
		\node [style=white-dot] (4) at (-0.5, -2.5) {$E$};
		\node [style=none] (5) at (0, 5.5) {\small $e_4$};
		\node [style=none] (6) at (3.5, 2) {\small $e_1$};
		\node [style=none] (7) at (-4.5, -4) {\color{red} \small $e_2$};
		\node [style=none] (8) at (0.5, -4) {\small $e_3+e_4$};
		\node [style=none] (9) at (-4.75, -0.5) {\color{red} \small $e_7+e_5$};
		\node [style=input] (10) at (-2, 2) {$A$};
		\node [style=none] (11) at (-5.25, 2) {\small $e_6$};
	\end{pgfonlayer}
	\begin{pgfonlayer}{edgelayer}
		\draw [style=single] (1) to (3);
		\draw [style=single] (1) to (4);
		\draw [style=single] (0) to (2);
		\draw [style=single] (0) to (10);
		\draw [style=single] (10) to (1);
	\end{pgfonlayer}
\end{tikzpicture}